\documentclass[useAMS,onecolumn]{mn2e}

\newif\ifAMStwofonts

\usepackage{graphicx}	% Including figure files
\usepackage{amssymb}	% Extra maths symbols
\usepackage{epsfig}
\usepackage{color}

\def\pmb#1{\mbox{\boldmath$#1$}}
\def\gtsim {>\kern-1.2em\lower1.1ex\hbox{$\sim$}}
\def\ltsim {<\kern-1.2em\lower1.1ex\hbox{$\sim$}}
\def\gtsim {>\kern-1.2em\lower1.1ex\hbox{$\sim$}}
\def\ltsim {<\kern-1.2em\lower1.1ex\hbox{$\sim$}}

\def\be{\begin{equation}}
\def\ee{\end{equation}}

\def\pmbmt#1{\pmb{\sf #1}}
\def\rmi{{\rm i}}

\begin{document}

\title[Axisymmetric spheroidal modes of magnetized stars]{Axisymmetric spheroidal modes of neutron stars magnetized with poloidal magnetic fields}

\author[U. Lee]{
Umin Lee$^{1}$\thanks{E-mail: lee@astr.tohoku.ac.jp},
\\
$^{1}$Astronomical Institute, Tohoku University, Sendai, Miyagi 980-8578, Japan\\
}

\date{Accepted XXX. Received YYY; in original form ZZZ}
\pubyear{2017}

\maketitle

\begin{abstract}
We calculate axisymmetric spheroidal modes of neutron stars magnetized with an axisymmetric 
poloidal magnetic field. 
We use polytropes of the indices $n\sim1$ as background equilibrium models of neutron stars where we ignore the 
deformation due to the magnetic fields.
For a poloidal magnetic field, axisymmetric normal modes of non-rotating stars are decoupled into spheroidal modes
and toroidal modes, and we can treat spheroidal modes separately from toroidal modes.
For the surface field strength $B_S$ ranging from $10^{14}$G to $10^{16}$G,
we calculate axisymmetric spheroidal magnetic modes whose oscillation frequency is proportional to $B_S$.
The typical oscillation frequency of the magnetic modes is $\sim 0.01\times\sqrt{GM/R^3}$ for $B_S\sim 10^{15}$G, where $M$ and $R$ are respectively the mass and radius of the star and $G$ is the gravitational constant.
For $M=1.4M_\odot$ and $R=10^6$cm, this frequency is $\sim 20$Hz, which may explain low frequency QPOs found for
SGR 1806-204 and SGR 1900+14.
We also find modes of frequency $\gtsim\sqrt{GM/R^3}$ corresponding to the radial fundamental and first harmonic modes.
No unstable magnetic modes are found for axisymmetric spheroidal oscillations of magnetized stars.
\end{abstract}

\section{Introduction}

Since the discovery of quasi-periodic-oscillations (QPOs) in the tail of the giant X/$\gamma$-ray flares of SGR 1806-204 (Israel et al 2005)  and SGR 1900+14 (e.g., Strohmayer \& Watts 2005, 2006, Watts \& Strohmayer 2006), 
oscillations of strongly magnetized neutron stars, i.e., magnetars (e.g., Woods \& Thompson 2006; Mereghetti 2008) have gained much attention in the astrophysics community.
Because of the suggestion by Duncan (1998), before the discovery of the magnetar QPOs,
that crustal torsional oscillations of neutron stars are most easily excited by seismic events, theoretical studies for the QPOs were first focused on torsional oscillations of the solid crust of neutron stars possessing strong magnetic fields and, later on, on global toroidal oscillations 
of magnetized neutron stars, whose frequencies 
are proportional to the strength of the magnetic fields.
We may call global modes of strongly magnetized stars Alfv\'en modes or magnetic modes when the oscillation
frequency is proportional to the field strength.
The frequency ranges of the crustal modes and the Alfv\'en modes are found consistent to
the observed QPO frequencies (e.g., Watts 2011).

Using a toy model for global oscillations of magnetized neutron stars, 
Levin (2006, 2007) suggested that torsional modes in the solid crust will be strongly damped by frequency resonance with Alfv\'en continuum in the fluid core.
Motivated by Levin's suggestions, many authors carried out MHD calculations that simulate time evolution of small amplitude axisymmetric toroidal perturbations to investigate the modal properties of magnetars 
(e.g., Sotani et al. 2008; Cerd\'a-Dur\'an et al. 2009; Colaiuda \& Kokkotas 2011; Gabler et al. 2011, 2012).
For example, Gabler et al (2011, 2012) showed that because of resonant damping 
with Alfv\'en continuum in the core the damping timescales of crustal torsional modes are too short to
be consistent with the observed QPOs, and suggested that Alfv\'en modes, instead of crustal modes, could be responsible for the QPOs if the magnetic fields are stronger than $\sim 10^{15}$G at the surface
since the damping timescales of Alfv\'en modes
are much longer than that of the crustal torsional modes that suffer resonant damping with Alfv\'en continuum.
Sotani \& Kokkotas (2009), on the other hand, computed axisymmetric $spheroidal$ oscillations of stars magnetized with a poloidal field, and they obtained Alfv\'en modes of low frequency as well as acoustic oscillations of high frequency.
They also suggested that Alfv\'en continuum in the core is irrelevant to the modal properties.

These early studies of oscillations of magnetars are mainly concerned with axisymmetric toroidal modes
of the stars, and pure toroidal modes do not produce any pressure and density perturbations.
For axisymmetric oscillations of magnetized stars, spheroidal and toroidal components of
the perturbed velocity fields are decoupled
and can be treated separately for a purely poloidal or toroidal magnetic field configuration if the effects of rotation are neglected.
This property remains true even for general relativistic treatment.
In other words, for non-axisymmetric oscillations of magnetized stars the spheroidal and toroidal 
velocity components are coupled even if we assume a pure poloidal or toroidal magnetic field and ignore the effects of rotation.

Lander et al. (2010), Passamonti \& Lander (2013) discussed, using MHD simulations, 
such non-axisymmetric oscillation modes of magnetized stars assuming a purely toroidal magnetic field.
For rotating magnetized stars, the modal property will be more complicated because spheroidal (polar) and toroidal (axial) components of the perturbed velocity fields are coupled and rotational modes such as
inertial modes and $r$-modes come in.
Lander \& Jones (2011) calculated non-axisymmetric oscillations of magnetized rotating stars for
a purely poloidal magnetic field. 
They obtained polar-led Alfv\'en modes which reduce to inertial modes in the limit of ${\cal M}/{\cal T}\to 0$, where $\cal M$ and $\cal T$ are magnetic and rotation energies of the star. 
Lander \& Jones (2011) also suggested that the axial-led Alfv\'en modes could be unstable.

For mixed poloidal and toroidal magnetic field configurations, coupled
spheroidal and toroidal velocity fields have to be considered to describe global oscillations of magnetized stars
even for axisymmetric oscillations.
Colaiuda \& Kokkotas (2012) calculated axisymmetric toroidal oscillations of
neutron stars adding a toroidal magnetic field to a poloidal one.
They found that the oscillation spectra of toroidal modes 
are significantly modified, losing their continuum character, by introducing a toroidal field component and
that the crustal torsional modes now become long-living oscillations. 
This finding may be similar to the finding by van Hoven \& Levin (2011, 2012) who suggested the existence of
discrete modes in the gaps between frequency continua, using a spectral method. 
Using MHD simulations, Gabler et al. (2013) discussed axisymmetric toroidal modes of magnetized stars
assuming various magnetic field configurations.
It is interesting to note that Gabler et al. (2013) found no long-lived discrete crustal modes in the gap
between Alfv\'en continua in the core, that is, their numerical results do not fully confirm
those by Colaiuda \& Kokkotas (2012) and by van Hoven \& Levin (2011, 2012).

Normal mode analysis is another method employed to study small amplitude oscillations of magnetized stars, 
where the time dependence of
oscillation is given by the factor $e^{\rmi \omega t}$ 
and we determine the oscillation frequency $\omega$ as an eigenfrequency of a system of linear differential equations
that govern the oscillations. 
Assuming a force-free dipole magnetic field, Lee (2007) computed axisymmetric normal modes of a neutron star.
Lee (2008) and Asai \& Lee (2014)
calculated $axisymmetric$ toroidal modes of magnetized neutron stars for poloidal magnetic fields, 
employing Newtonian gravity in the former and general relativity in the latter, while
Asai, Lee, \& Yoshida (2016) calculated $non$-$axisymmetric$ magnetic modes for a poloidal magnetic field
and obtained both stable and unstable magnetic modes.
Assuming a toroidal magnetic field, on the other hand, Asai, Lee, \& Yoshida (2015) computed 
non-axisymmetric oscillation modes of rotating stars, taking account of the equilibrium deformation 
caused by the toroidal magnetic field.
They obtained $g$-, $f$-, $p$-modes for non-rotating stars and $r$-modes and inertial modes for rotating stars.
It is important to note that the numerical results obtained by the normal mode analyses are not necessarily consistent with those obtained by numerical MHD simulations.
The frequency ranges of magnetic modes obtained by normal mode calculations and MHD simulations are similar
between the two methods of analysis. 
However, although Lee (2008) and Asai \& Lee (2014) obtained discrete toroidal magnetic normal modes of neutron stars for a poloidal magnetic field, MHD simulations for axisymmetric toroidal oscillations do not necessarily support the existence of such discrete magnetic normal modes.
The reason for the discrepancy, however, is not necessarily well understood.

We may employ normal mode analysis to investigate the stability of magnetized stars, but its application to
general field configurations is extremely difficult to formulate.
In this respect the stability analysis based on variational principle will be more flexible and tractable.
See Herbric \& Kokkotas (2017) and Akg\"un et al (2013) for recent studies, and see also Tayler (1973), Markey \& Tayler (1973, 1974) for classical works.

In this paper, we revisit the problem of axisymmetric spheroidal normal modes of magnetized stars for a poloidal
magnetic field.
Although we employ Newtonian gravity, this paper may be regarded as a normal mode analysis version of Sotani \& Kokkotas (2009), who carried out
numerical MHD simulations for small amplitude perturbations in general relativistic framework.
\S 2 is for method of solution and \S 3 is for numerical results.
We conclude in \S 4.

\section{Method of solution}

\subsection{Equilibrium State}

For a static magnetic field, we assume an axisymmetric poloidal field given in spherical polar coordinates $(r,\theta,\phi)$ by
\be
B_r=2f(r)\cos\theta, \quad B_\theta=-{1\over r}{dr^2f(r)\over dr}\sin\theta, \quad B_\phi=0,
\ee
for which $\nabla\cdot\pmb{B}=0$.
If we assume that the electric current is toroidal and is given by $\pmb{J}_\phi=cj_\phi r\sin\theta\pmb{e}_\phi$
with $c$ being the velocity of light, the Ampere law given by
$
\nabla\times\pmb{B}=4\pi \pmb{J}_\phi/c
$
leads to
\be
{d^2f\over dr^2}+{4\over r}{df\over dr}=-4\pi j_\phi,
\ee
where we assume that in the vicinity of $r=0$ the function $f$ behaves as
\be
f(r)=\alpha_0+O(r^2)
\ee
with $\alpha_0$ being a constant.
The Lorenz force is then given by
\be
(\nabla\times\pmb{B})\times\pmb{B}
=4\pi j_\phi\nabla\left(r^2f\sin^2\theta\right),
\ee 
and the hydrostatic equation reduces to
\be
{1\over\rho}\nabla p+\nabla\Phi-{j_\phi\over\rho}\nabla\left(r^2f\sin^2\theta\right)=0,
\ee
where $p$ is the pressure, $\rho$ is the mass density, and $\Phi$ is the gravitational potential.
This may suggest that a possible choice of the function $j_\phi$ is
\be
j_\phi=c_0\rho,
\ee
where $c_0$ is a constant.
For this choice of $j_\phi$ the hydrostatic equation simplifies to
\be
{1\over\rho}\nabla p+\nabla\left(\Phi-c_0r^2f\sin^2\theta\right)=0,
\ee
where the term $c_0r^2f\sin^2\theta$ is responsible for deviation of the equilibrium from spherical symmetry.
For simplicity, however, we ignore the term $c_0r^2f\sin^2\theta$ to construct equilibrium structures
so that $p$, $\rho$, and $\Phi$ depend only on the radial distance $r$ from the centre.
See Asai et al (2016),
who took account of the equilibrium deformation caused by toroidal magnetic fields to calculate
various oscillation modes.

The constants $c_0$ and $\alpha_0$ are determined by using surface boundary conditions for 
the function $f$.
We assume that $j_\phi=0$ outside the star, and hence the exterior solution $f^{\rm ex}$
is given by $f^{\rm ex}=\mu_b/r^3$, where $\mu_b$ is the magnetic dipole moment of the star.
The constants $c_0$ and $\alpha_0$ are determined so that the interior solution $f$ and $df/dr$
are matched with the exterior solution $f^{\rm ex}$ and $df^{\rm ex}/dr$ at the surface $r=R$
with $R$ being the radius of the star.

\begin{figure}
\begin{center}
\resizebox{0.33\columnwidth}{!}{
\includegraphics{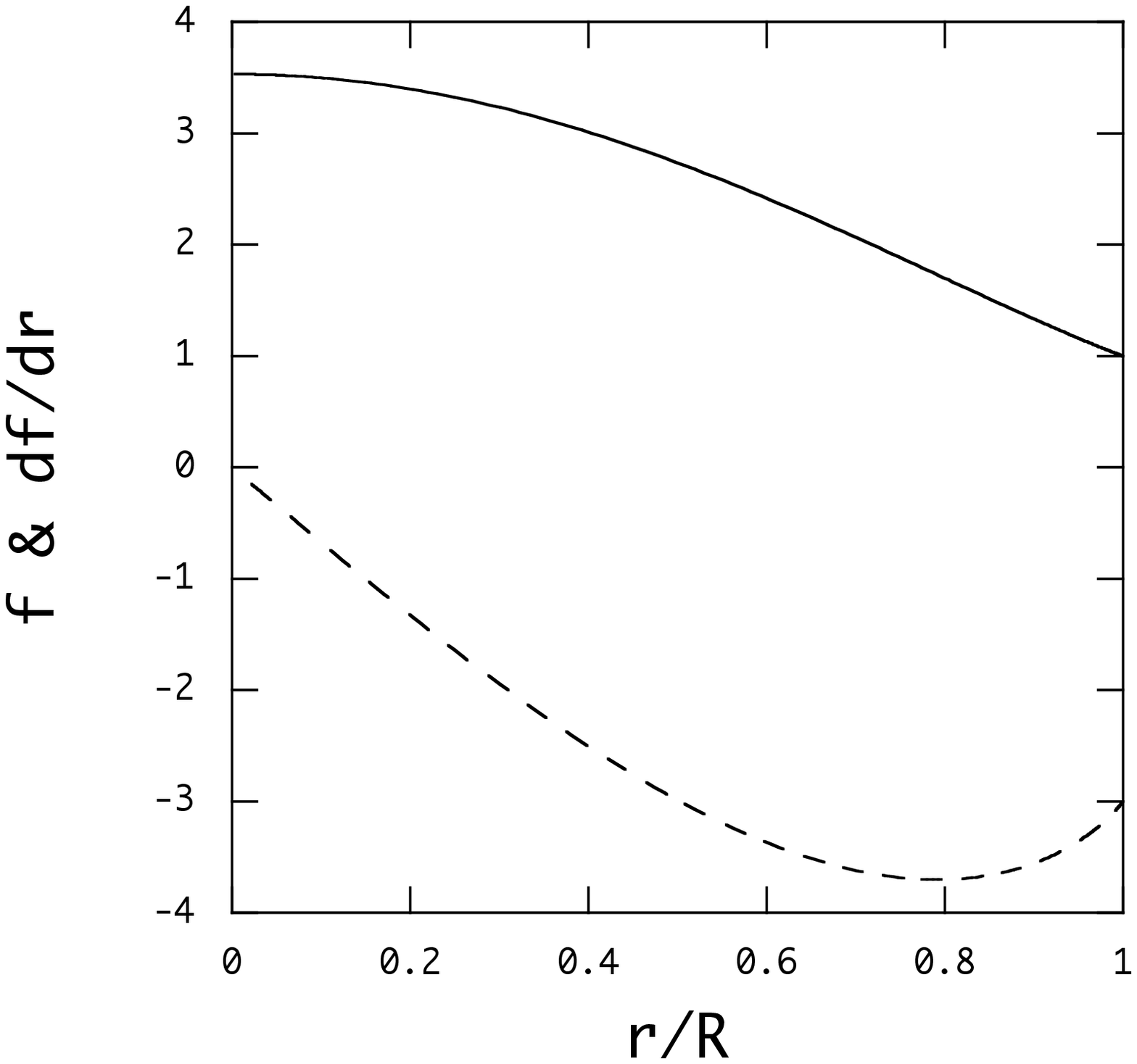}}
\resizebox{0.33\columnwidth}{!}{
\includegraphics{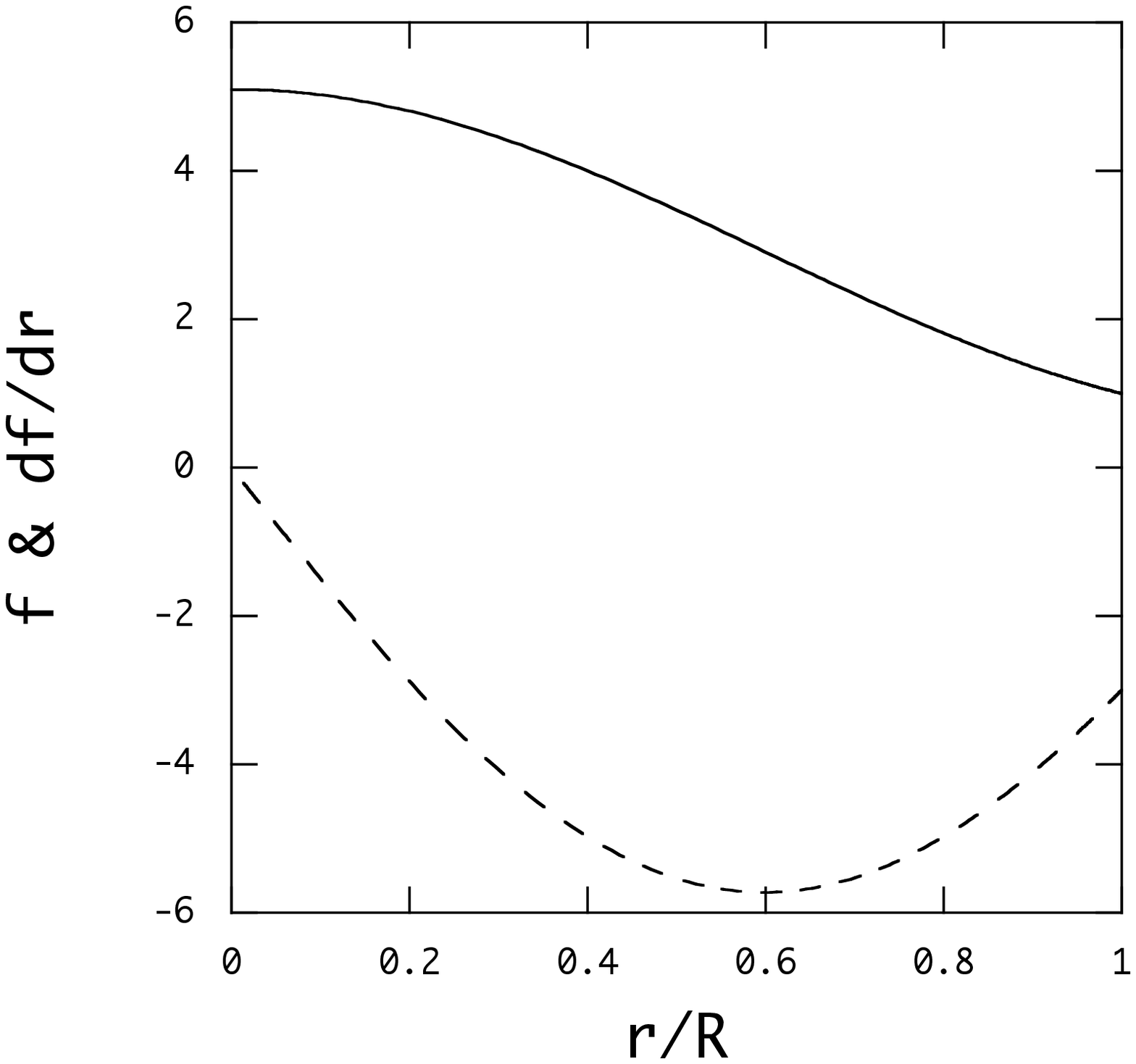}}
\resizebox{0.33\columnwidth}{!}{
\includegraphics{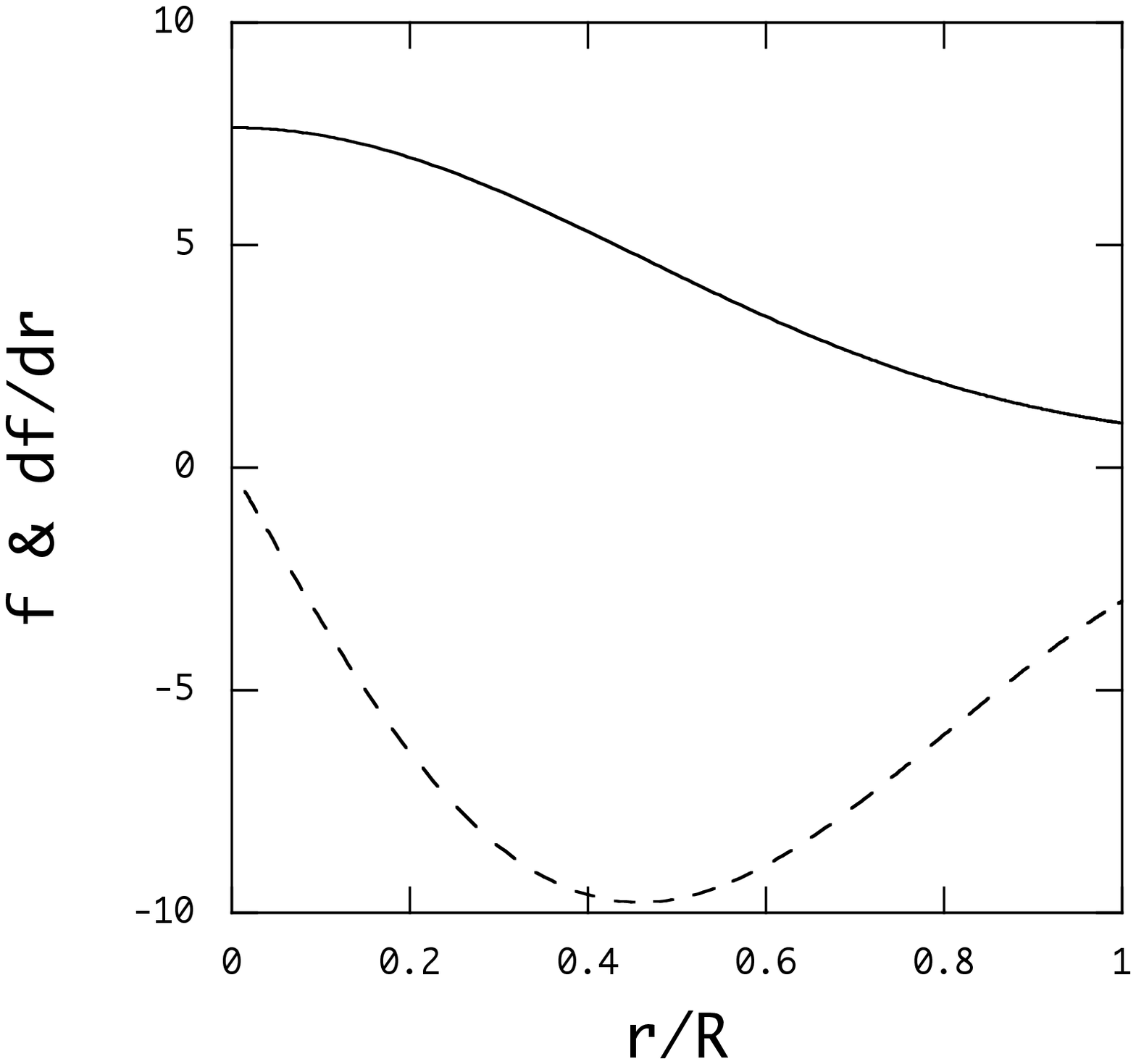}}
\end{center}
\caption{Functions $f$ (solid line) and $df/dr$ (dashed line) versus $x=r/R$ for polytropes of the indices $n=0.5$, 1, and 1.5, from the left to right panels, where $f$ and $df/dr$ normalized respectively by $B_S\equiv \mu_b/R^3$ and $B_S/R$ are shown.
}
\label{fig:fdfdr}
\end{figure}

\begin{figure}
\begin{center}
\resizebox{0.33\columnwidth}{!}{
\includegraphics{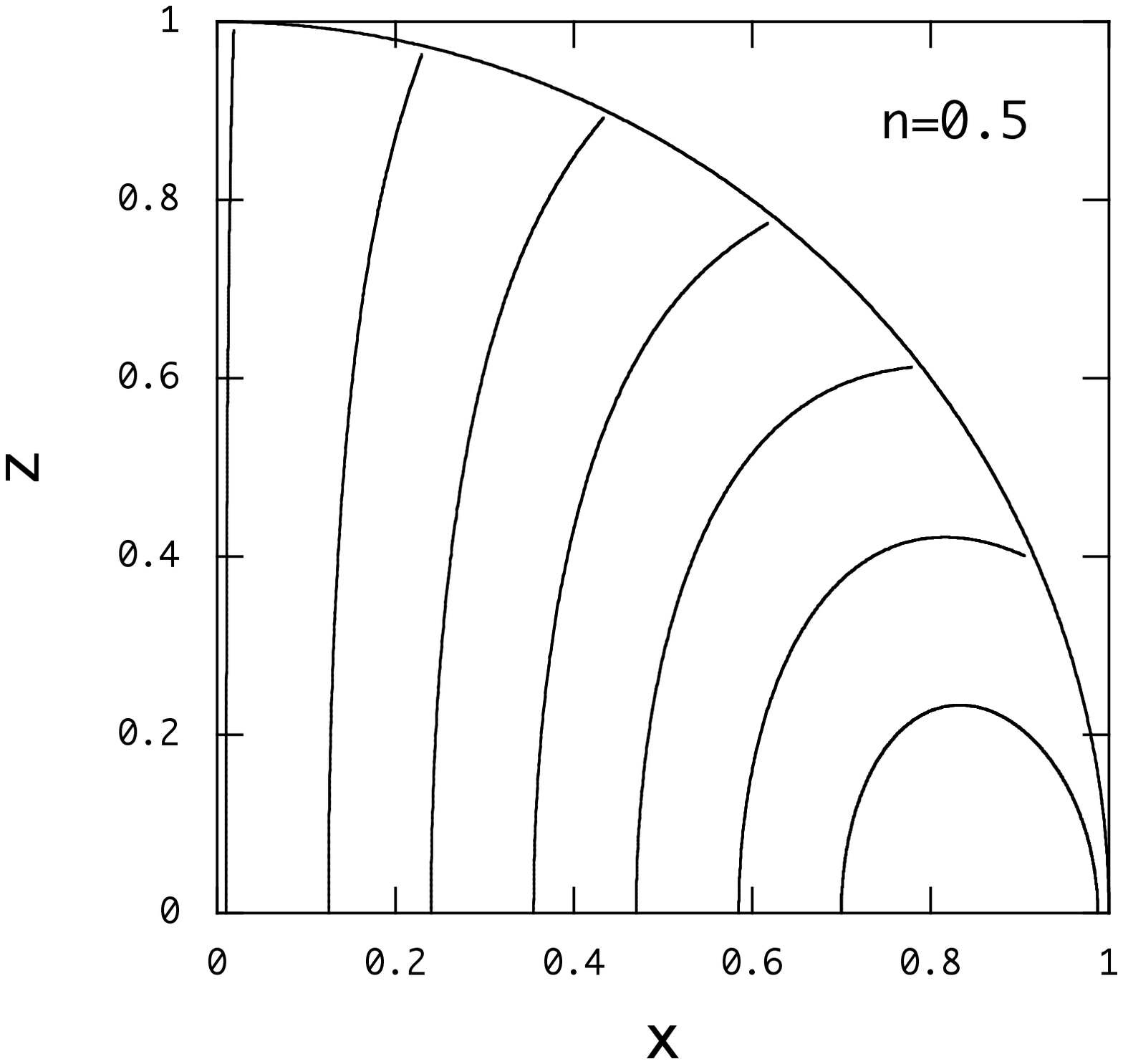}}
\resizebox{0.33\columnwidth}{!}{
\includegraphics{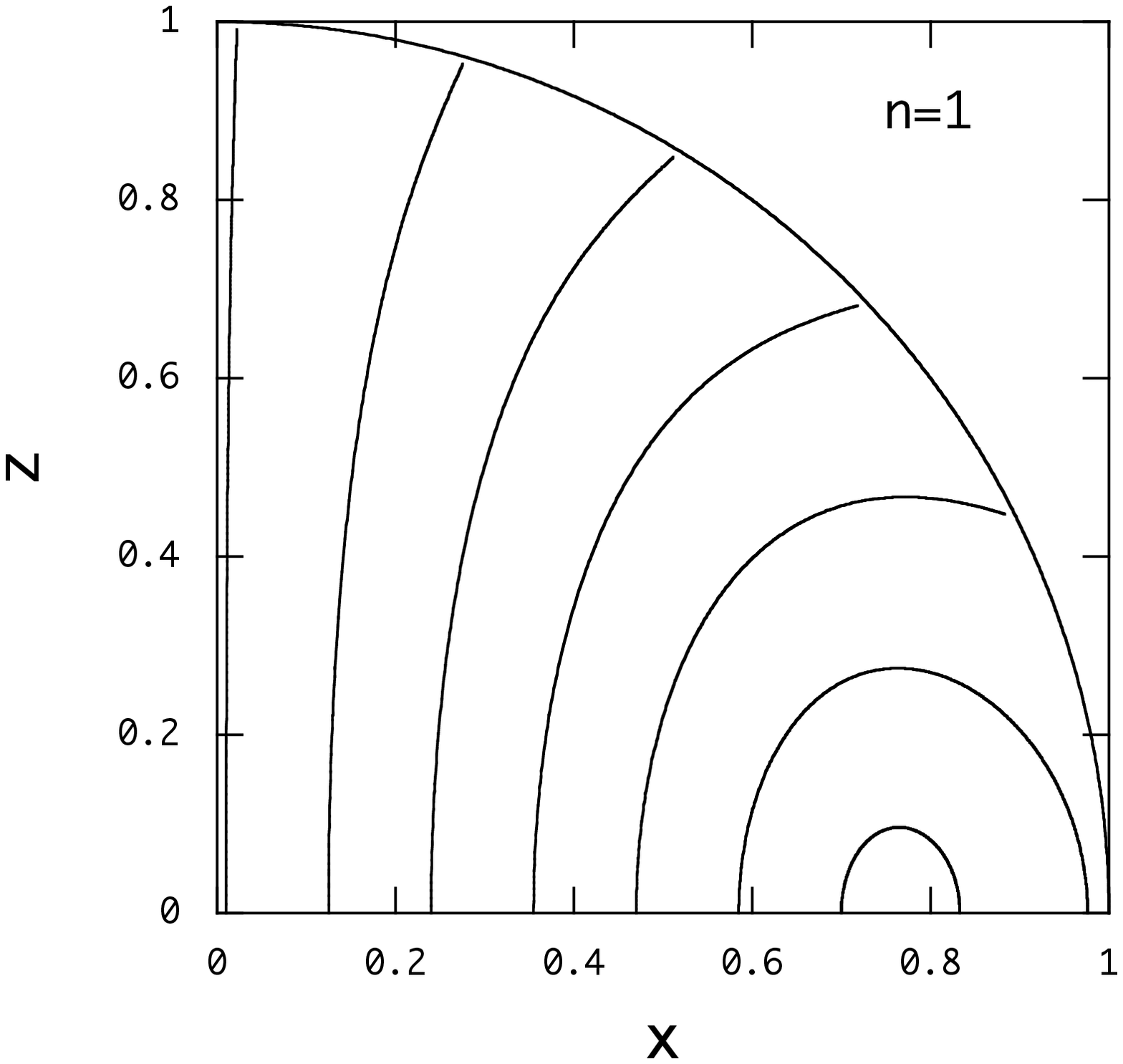}}
\resizebox{0.33\columnwidth}{!}{
\includegraphics{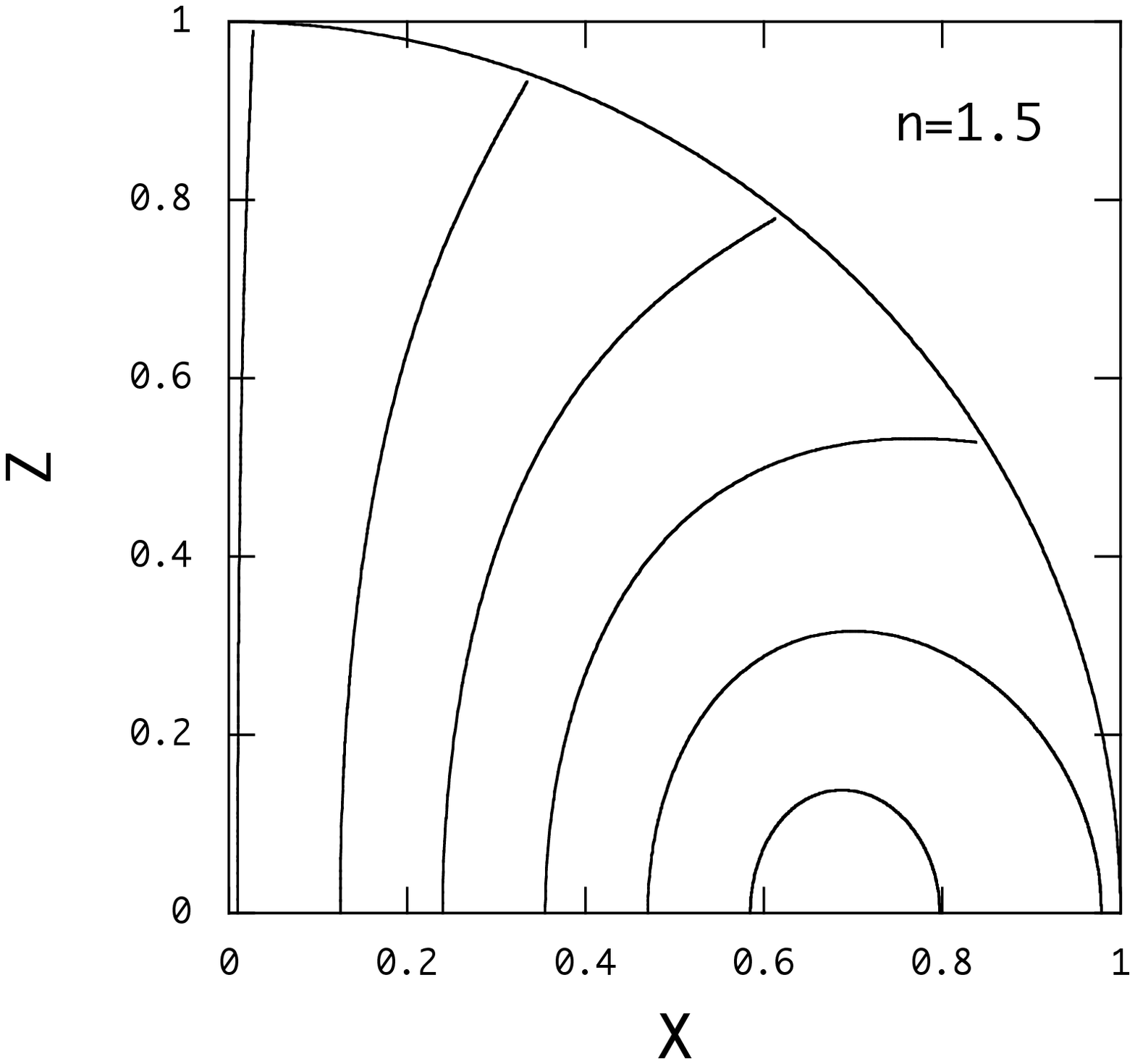}}
\end{center}
\caption{Internal magnetic field lines in the x-z plane for polytropes of the indices $n=0.5$, 1, and 1.5.
}
\label{fig:bfldlines}
\end{figure}

We use polytropes of the indices $n=0.5$, 1, and 1.5 as background models for modal analysis in this paper,
where no deformation of the polytropes is considered, and the mass $M=1.4M_\odot$ and the radius $R=10^6$cm are assumed.
As the index $n$ increases, the concentration of the mass density into the central region becomes stronger.
Figures \ref{fig:fdfdr} and \ref{fig:bfldlines} show the functions $f$ and $df/dr$ versus $x=r/R$ and
the internal magnetic field lines on the $x$-$z$ plane for the polytropes, where $f$ and $df/dr$ normalized respectively by $B_S\equiv\mu_b/R^3$ and $B_S/R$ are shown, and  
the $x$ axis is in the equatorial plane and the $z$ axis is along the magnetic axis of the star.
With increasing $n$ the centre of the closed magnetic field region moves inward and the volume of the region increases.

\subsection{Perturbation Equations}

Neglecting equilibrium deformation caused by the poloidal magnetic field and
assuming spherical symmetry of the star,
we linearize ideal MHD equations to derive the oscillation equations of magnetized stars, 
which are
\be
\rho'+\nabla\cdot\left(\rho\pmb{\xi}\right)=0,
\label{eq:conteq}
\ee
\be
%-\sigma^2\pmb{\xi}
-\omega^2\pmb{\xi}
+{1\over\rho}\nabla p'-{\rho'\over\rho^2}{dp\over dr}\pmb{e}_r
-{1\over 4\pi \rho}\left[\left(\nabla\times\pmb{B}'\right)\times\pmb{B}+\left(\nabla\times\pmb{B}\right)\times
\pmb{B}'\right]=0,
\label{eq:eqmot}
\ee
\be
\pmb{B}'=\nabla\times\left(\pmb{\xi}\times\pmb{B}\right),
\label{eq:inductioneq}
\ee
\be
{\rho'\over\rho}={V\over\Gamma_1}{p'\over\rho g r}-rA{\xi_r\over r},
%{\rho'\over\rho}={p'\over\rho g r}-rA{\xi_r\over r},
\label{eq:rhoprime}
\ee
where $\pmb{\xi}$ is the displacement vector and
the prime $(')$ indicates Euler perturbations, and we have assumed that the time dependence of the 
perturbations is given by the factor $e^{\rmi \omega t}$ with $\omega$ being the oscillation frequency.
The Schwarzschild discriminant $rA$ may be defined as
\be
rA={d\ln\rho\over d\ln r}-{1\over\Gamma_1}{d\ln p\over d\ln r},
\ee
where $\Gamma_1=(\partial\ln p/\partial\ln \rho)_{\rm ad}$.
Note that we have applied the Cowling approximation neglecting $\Phi'$, the Eulerian perturbation of
the gravitational potential, and that the linearized induction equation (\ref{eq:inductioneq}) guarantees 
$
\nabla\cdot\pmb{B}'=0.
$
For polytropes of the index $n$, the adiabatic exponent for the perturbations are assumed
to be given by
\be
{1\over\Gamma_1}={n\over n+1}+\gamma,
\ee
where $\gamma$ is a constant, and 
we have $rA=-\gamma(d\ln p/d\ln r)\equiv\gamma V$.
The equilibrium configuration may be called radiative for $\gamma<0$, isentropic for $\gamma=0$, and convective for $\gamma>0$.

Because of the Lorentz terms in equation (\ref{eq:eqmot}), separation of variables is not possible
to represent the perturbations of magnetized stars.
We therefore employ a series expansion for the perturbations.
For the displacement vector $\pmb{\xi}$, we write for axisymmetric perturbations of $m=0$
\be
{\xi_r\over r}=\sum_j^{j_{\rm max}}S_{l_j}(r)Y_{l_j}^m(\theta,\phi)e^{\rmi\omega t},
\ee
\be
{\xi_\theta\over r}=\sum_j^{j_{\rm max}}
H_{l_j}(r){\partial\over\partial\theta}Y_{l_j}^m(\theta,\phi)
e^{\rmi\omega t},
\ee
\be
{\xi_\phi\over r}=-\sum_j^{j_{\rm max}}
T_{l'_j}(r){\partial\over\partial \theta}Y_{l'_j}^m(\theta,\phi)
e^{\rmi\omega t},
\ee
and for the perturbed magnetic fields $\pmb{B}'$
\be
{B'_r\over f}=\sum_j^{j_{\rm max}}b^S_{l'_j}(r)Y_{l'_j}^m(\theta,\phi)e^{\rmi\omega t},
\ee
\be
{B'_\theta\over f}=\sum_j^{j_{\rm max}}
b^H_{l'_j}(r){\partial\over\partial\theta}Y_{l'_j}^m(\theta,\phi)
e^{\rmi\omega t},
\ee
\be
{B'_\phi\over f}=-\sum_j^{j_{\rm max}}b^T_{l_j}(r){\partial\over\partial \theta}Y_{l_j}^m(\theta,\phi)
e^{\rmi\omega t},
\ee
where $l_j=2(j-1)$ and $l'_j=2j-1$ for even modes and
$l_j=2j-1$ and $l'_j=2(j-1)$ for odd modes for $j=1,~ 2,~3,\cdots$.
The pressure perturbation $p'_{l_j}$ is given by
\be
p'=\sum_j^{j_{\rm max}}p'_{l_j}(r)Y_{l_j}^m(\theta,\phi)e^{\rmi\omega t}.
\ee

Substituting the series expansion into the perturbed basic equations (\ref{eq:conteq}) to (\ref{eq:rhoprime}),
we obtain sets of linear ordinary differential equations for the expansion coefficients.
For the dependent variables defined as
\be
\pmb{y}_1=\left({S_{l_j}}\right), \quad \pmb{y}_2=\left({p'_{l_j}\over \rho g r}\right), \quad
\pmb{y}_3=\left(H_{l_j}\right), \quad \pmb{y}_4=\left(b^H_{l'_j}\right),\quad \pmb{b}^S=\left(b^S_{l'_j}\right),
\quad \pmb{t}=\left(T_{l'_j}\right), \quad \pmb{b}^T=\left(b^T_{l_j}\right),
\ee
we obtain for the spheroidal components $\pmb{y}_i$ with $i=1,~2,~3,~4$
\be
r{d\pmb{y}_1\over dr}=\left({V\over \Gamma_1}-3\right)\pmb{y}_1-{V\over\Gamma_1}\pmb{y}_2+\pmbmt{\Lambda}_0\pmb{y}_3,
\label{eq:y1}
\ee
\begin{eqnarray}
r{d\pmb{y}_2\over dr}&=&\left[\left(c_1\bar\omega^2+rA\right)\pmbmt{1}-{1\over 8\pi V}{f^2\over p}\left({{\cal Q}\over f}\right)^2
{{\cal R} \over f}\pmbmt{C}_0\pmbmt{B}_0^{-1}\pmbmt{W}_0\pmbmt{W}_1\right]\pmb{y}_1
-\left[\left(rA+U-1\right)\pmbmt{1}+{1\over 2}{{\cal Q}\over f}\pmbmt{C}_0\pmbmt{B}_1^{-1}\pmbmt{\Lambda}_0\right]\pmb{y}_2\nonumber\\
&&+{{\cal Q}\over f}\left({1\over 2}c_1\bar\omega^2\pmbmt{C}_0\pmbmt{B}_1^{-1}\pmbmt{\Lambda}_0
-{1\over4\pi V}{f^2\over p}{{\cal R}\over f}\pmbmt{C}_0\pmbmt{B}_1^{-1}\pmbmt{W}_0\pmbmt{B}_0\right)\pmb{y}_3-{1\over 4\pi V}{f^2\over p}{{\cal R}\over f}\pmbmt{C}_0\pmb{y}_4,
\label{eq:y2}
\end{eqnarray}
\begin{eqnarray}
r{d\pmb{y}_3\over dr}={1\over 2}{{\cal Q}\over f}\left(3-{V\over\Gamma_1}+{{\cal Q}\over f}-{d\ln {\cal Q}/f\over d\ln r}\right)
\pmbmt{B}_0^{-1}\pmbmt{W}_1\pmb{y}_1+{1\over 2}{V\over\Gamma_1}{{\cal Q}\over f}\pmbmt{B}_0^{-1}\pmbmt{W}_1\pmb{y}_2
+{{\cal Q}\over f}\left(\pmbmt{1}-{1\over 2}\pmbmt{B}_0^{-1}\pmbmt{W}_1\pmbmt{\Lambda}_0\right)\pmb{y}_3
+{1\over 2}\pmbmt{B}_0^{-1}\pmbmt{\Lambda}_1\pmb{y}_4,
%+{1\over 2}\pmbmt{B}_0^{-1}\pmbmt{\Lambda}_0\pmb{y}_4,
\label{eq:y3}
\end{eqnarray}
\begin{eqnarray}
r{d\pmb{y}_4\over dr}
%&=&-2\pi V{p\over f^2}\pmbmt{B}_1^{-1}\pmbmt{\Lambda}_0\left(c_1\bar\omega^2\pmb{y}_3
%-\pmb{y}_2\right)-{d\ln rf\over d\ln r}\pmb{y}_4+\left(\pmbmt{1}+
%{1\over 2}{{\cal R}\over f}\pmbmt{B}_1^{-1}\pmbmt{W}_0\right)\pmb{b}^S\nonumber\\
&=&{{\cal Q}\over f}\left(\pmbmt{1}+{1\over 2}{{\cal R}\over f}\pmbmt{B}_1^{-1}\pmbmt{W}_0\right)\pmbmt{W}_1\pmb{y}_1
+2\pi V{p\over f^2}\pmbmt{B}_1^{-1}\pmbmt{\Lambda}_0\pmb{y}_2
+\left[2\left(\pmbmt{1}+{1\over 2}{{\cal R}\over f}\pmbmt{B}_1^{-1}\pmbmt{W}_0\right)\pmbmt{B}_0-2\pi V{p\over f^2}c_1\bar\omega^2\pmbmt{B}_1^{-1}\pmbmt{\Lambda}_0\right]\pmb{y}_3\nonumber\\
&&-{d\ln rf\over d\ln r}\pmb{y}_4,
\label{eq:y4}
\end{eqnarray}
and for the toroidal components $\rmi\pmb{t}$ and $\rmi\pmb{b}^T$
\be
r{d\rmi\pmb{t}\over dr}=
-{{\cal Q}\over f}\left({1\over 2}\pmbmt{B}_1^{-1}\pmbmt{\Lambda}_0\pmbmt{C}_0-\pmbmt{1}\right)\rmi\pmb{t}
+{1\over 2}\pmbmt{B}_1^{-1}\pmbmt{\Lambda}_0\rmi\pmb{b}^T,
\ee
\be
r{d\rmi \pmb{b}^T\over dr}=-2\pi V{p\over f^2}c_1\bar\omega^2\pmbmt{B}_0^{-1}\pmbmt{\Lambda}_1\rmi\pmb{t}
-\left({1\over 2}{{\cal Q}\over f}\pmbmt{B}_0^{-1}\pmbmt{W}_1\pmbmt{\Lambda}_0
+{d\ln fr\over d\ln r}\pmbmt{1}\right)\rmi\pmb{b}^T,
\label{eq:bt}
\ee
where
\be
\bar\omega={\omega\over\sigma_0}, \quad \sigma_0=\sqrt{GM\over R^3}, \quad 
V=-{d\ln p\over d\ln r}, \quad U={d\ln M_r\over d\ln r}, \quad c_1={(r/R)^3\over M_r/M},
\ee
\be
{\cal Q}=-{1\over r}{d(r^2f)\over dr},
\quad {\cal R}
=-r^2\left({d^2f\over dr^2}+{4\over r}{df\over dr}\right),
\ee
and $\pmbmt{1}$ denotes the identity matrix, and 
the matrices $\pmbmt{B}_0$, $\pmbmt{B}_1$, $\pmbmt{W}_0$, and $\pmbmt{W}_1$ are defined as
\be
\pmbmt{B}_0=\pmbmt{Q}_1\pmbmt{\Lambda}_0+\pmbmt{C}_1, \quad \pmbmt{B}_1=\pmbmt{Q}_0\pmbmt{\Lambda}_1+\pmbmt{C}_0, \quad \pmbmt{W}_0=2\pmbmt{Q}_0+\pmbmt{C}_0, \quad \pmbmt{W}_1=2\pmbmt{Q}_1+\pmbmt{C}_1,
\ee
and the definition of the matrices $\pmbmt{Q}_0$, $\pmbmt{Q}_1$, $\pmbmt{C}_0$, $\pmbmt{C}_1$, $\pmbmt{\Lambda}_0$, and $\pmbmt{\Lambda}_1$ is given in Yoshida \& Lee (2000).
It is important to note that since the expansion coefficient $H_{l_1}=H_0$, for example, vanishes identically for even modes, 
we take $H_{l_2}$ to $H_{l_{j_{{\rm max}}+1}}$ as dependent variables and hence we have to redefine the matrices given above accordingly (see, e.g., Lee 2008).
To simplify the set of differential equations for the spheroidal components we have used
\be
\pmb{b}^S={Q\over f}\pmbmt{W}_1\pmb{y}_1+2\pmbmt{B}_0\pmb{y}_3,
\label{eq:induction_r}
\ee
which comes from the radial component of the induction equation (\ref{eq:inductioneq}), and
\be
r{d\pmb{b}^S\over dr}={Q\over f}\pmb{b}^S+\pmbmt{\Lambda}_1\pmb{y}_4,
\label{eq:nablab}
\ee
which comes from $\nabla\cdot\pmb{B}'=0$.
As shown by equations (\ref{eq:y1}) to (\ref{eq:bt}), for axisymmetric modes of $m=0$
the sets of differential equations for the spheroidal and toroidal components are decoupled from each other
(see, e.g., Lee 2007).

In this paper, we discuss axisymmetric ($m=0$) spheroidal modes of neutron stars magnetized by a poloidal magnetic field.
We solve the set of linear ordinary differential equations from (\ref{eq:y1}) to (\ref{eq:y4}), using
a relaxation method, as an eigenvalue problem of the frequency $\omega$, applying
boundary conditions at the center and the surface of the star.
The inner boundary conditions we use are the regularity conditions for the perturbations
$\pmb{\xi}$, $\pmb{B}'$, and $p'$ at the stellar centre.
The outer boundary conditions are $\delta p=0$ and $\pmb{b}^S+\pmbmt{L}^+\pmb{b}^H=0$ at the surface of the star where $\delta$ indicates
the Lagrangian perturbation and $(\pmbmt{L}^+)_{ij}=(l'_j+1)\delta_{ij}$.
See Asai, Lee, \& Yoshida (2016) for the details of the surface boundary conditions.

\begin{table*}
\begin{center}
\caption{Normalized eigenfrequency $\bar{\omega}$ of the magnetic modes of polytropes of the indices $n=0.5$, 1, and 1.5 for $\gamma=0$.}
\begin{tabular}{cccccc}
\hline
 & even  &  & & odd &\\
\hline
$n=0.5$ & $n=1$ & $n=1.5$ & $n=0.5$ & $n=1$ & $n=1.5$ \\
\hline
\multicolumn{6}{c}{$B_S=10^{14}$G} \\
\hline
0.001154 & 0.001425  & 0.001833  & 0.0005510 & 0.0007268  & 0.001006\\
0.001134 & 0.001401  & $\cdots$  & 0.0005001 & 0.0006937  & 0.0009860\\
$\cdots$ & $\cdots$ & $\cdots$  & 0.0004848 & 0.0006794 & $\cdots$\\
\hline
\multicolumn{6}{c}{$B_S=10^{15}$G} \\
\hline
0.01154 & 0.01425  & 0.01835  & 0.005510 & 0.007268  & 0.01001\\
0.01134 & 0.01400  & 0.01816  & 0.005001 & 0.006939  & 0.009878\\
$\cdots$ & $\cdots$ & $\cdots$ & 0.004851 & 0.006831 & 0.009756\\
\hline
\multicolumn{6}{c}{$B_S=10^{16}$G} \\
\hline
0.1153 & 0.1422  & 0.1828  & 0.05502 & 0.07253  & 0.1003\\
0.1134 & 0.1405  & 0.1809  & 0.04993 & 0.06924  & 0.09864\\
0.1122 & $\cdots$ & $\cdots$ & 0.04845 & 0.06841 & 0.09744\\
\hline
\hline
\end{tabular}
\medskip
\end{center}
\end{table*}

\section{Numerical Results}

We use polytropes of the indices $n=0.5$, 1, and 1.5 as equilibrium models for modal analyses, 
where the mass $M=1.4M_\odot$ and the radius $R=10^6$cm are assumed.
For a given field strength $B_S\equiv \mu_b/R^3$ and a given expansion length $j_{\rm max}$,
we find numerous solutions to the set of equations (\ref{eq:y1}) to (\ref{eq:y4}),
and we pick up those solutions whose eigenfrequency $\omega$ and eigenfunctions converge 
with increasing $j_{\rm max}$.
For a given field strength $B_S$, we usually find only a few solutions that reach a good convergence.

\subsection{Magnetic Modes}

Numerical results for axisymmetric spheroidal magnetic modes
are summarized in Table 1 where the normalized eigenfrequency $\bar\omega=\omega/\sigma_0$
with $\sigma_0=\sqrt{GM/R^3}$
is tabulated for both even and odd modes for the field strength $B_S=10^{14}$G, $10^{15}$G, and $10^{16}$G
and for $\gamma=0$.
We may convert the normalized frequency $\bar\omega$ to $\nu$(Hz) using $\nu\approx 2170\times\bar\omega$ for
$M=1.4M_\odot$ and $R=10^6$cm.
The magnetic modes in the table may correspond to polar Alfv\'en modes discussed by Sotani \& Kokkotas (2009).
Note that only magnetic modes that attain a good convergence with increasing $j_{\rm max}$ are tabulated. 
As in the case of non-axisymmetric ($m\not=0$) spheroidal magnetic modes of polytropes, 
the frequency $\bar\omega$ of the magnetic modes decreases as the number of nodes of the dominating eigenfunctions increases (see Asai et al 2016).
From the table we find that the frequency of the magnetic modes is in a good approximation 
proportional to the field strength $B_S$, which may suggest that compressibility as the restoring force for acoustic modes plays only a minor role for low frequency magnetic modes.
We find it difficult to compute magnetic modes for field strength $B_S\ltsim 10^{13}$G even for
$\gamma=0$.
This situation is different from that for axisymmetric toroidal magnetic modes of polytropes for
poloidal magnetic fields, for which we can find toroidal magnetic modes for any values of $B_S$ (e.g., Lee 2008).

Figures 3 shows the wave patterns 
$\xi_r(x,z)$ and $\xi_\theta(x,z)$ of axisymmetric magnetic modes of even parity for
polytropes of the indices $n=0.5$, 1, and 1.5 for $B_S=10^{15}$G, where $x=r\sin\theta$ and $z=r\cos\theta$,
and the amplitudes in each panel are normalized by their maximum value.
The normalized frequency of the modes in the figure is 
$\bar\omega=0.01154$, 0.01425, and 0.01835 for $n=0.5$, 1, and 1.5, respectively.
The magnetic modes in the figure are those of highest frequency $\bar\omega$ for given $n$ and $B_S$ and 
show simplest wave patterns.
For even modes, the patterns of $\xi_r$ is symmetric about the equator and those of $\xi_\theta$ antisymmetric.
The amplitudes of $\xi_r$ is confined to the region around the magnetic axis, and this amplitude
confinement becomes stronger as $n$ increases.
The same is true for $\xi_\theta$, which show some complex structures in the region of closed magnetic field lines (see Figure \ref{fig:bfldlines}).
This complex patterns of $\xi_\theta(x,z)$ in the region of closed field lines become
more conspicuous as $n$ increases.

Figure 4 shows the wave patterns $\xi_r(x,z)$ and $\xi_\theta(x,z)$ of the magnetic mode of even parity
having the frequency $\bar\omega=0.01400$ for $n=1$ and $B_S=10^{15}$G.
Comparing to the middle panels of Figure 3, we note that the numbers of nodal lines parallel to the magnetic axis
in the patterns of $\xi_r(x,z)$ has increased by one.
As the number of nodal lines increases with decreasing frequency $\bar\omega$,
the patterns tend to show complex structures around and within the region of closed field lines,
which may make it difficult to compute correctly convergent magnetic modes.

Figure 5 shows the wave patterns of $\xi_r(x,z)$ and $\xi_\theta(x,z)$ for the odd parity magnetic modes of highest frequency for $B_S=10^{15}$G
and their frequencies are $\bar\omega=0.00551$, 0.007268, and 0.01001 for
$n=0.5$, 1, and 1.5, respectively.
For odd modes, the patterns of $\xi_r$ and $\xi_\theta$ are respectively antisymmetric and symmetric about the
equator.
The amplitudes are also confined to the region along the magnetic axis and the region of closed field lines
does not necessarily manifest itself conspicuously. 
Comparing Figures 3 and 5, we find that the wave patterns $\xi_r$ and $\xi_\theta$ of the odd magnetic modes of highest frequency are much less complicated than those of the even magnetic modes of highest frequency.

\begin{figure}
\begin{center}
\resizebox{0.33\columnwidth}{!}{
\includegraphics{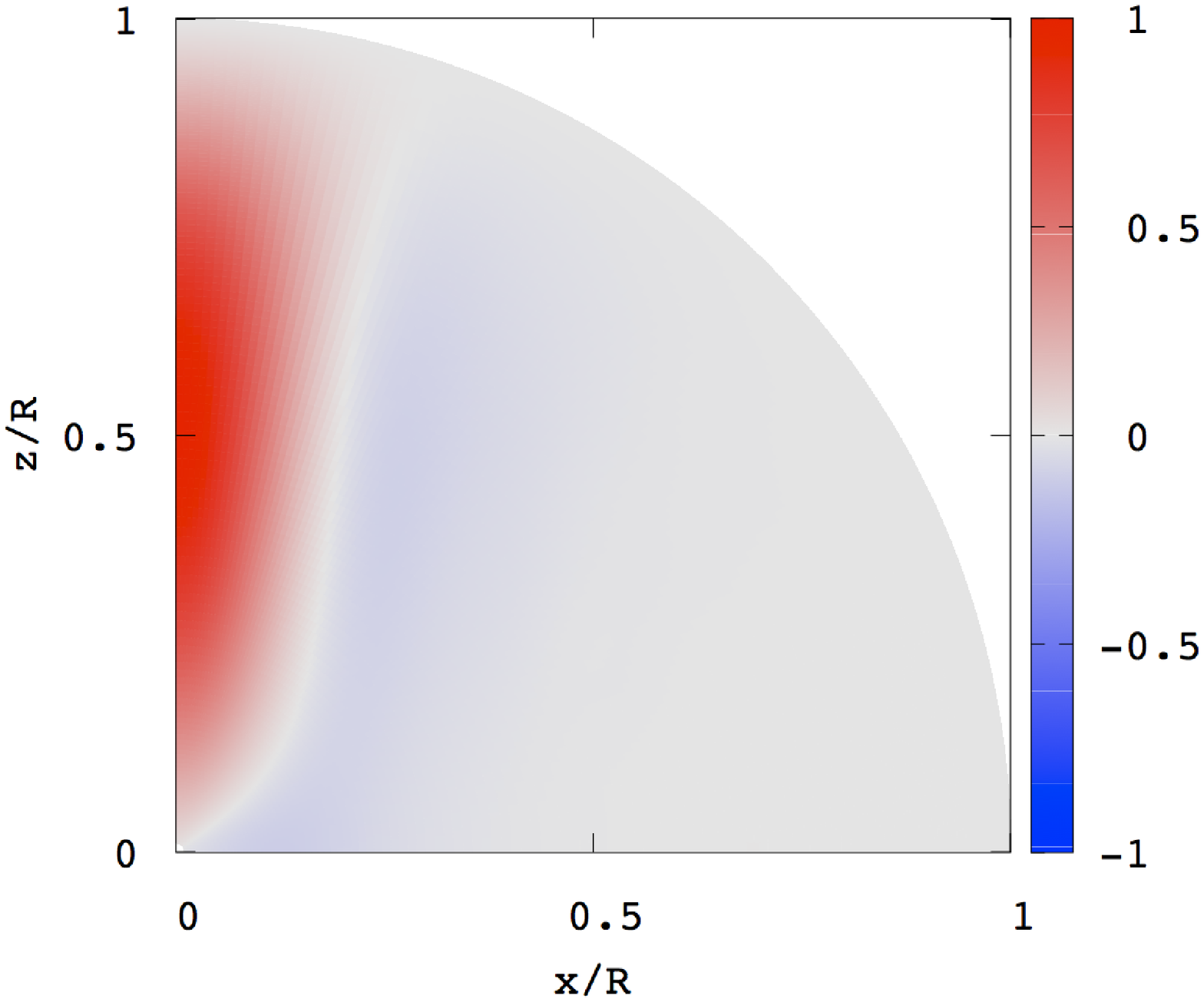}}
\hspace*{-1.41cm}
\resizebox{0.33\columnwidth}{!}{
\includegraphics{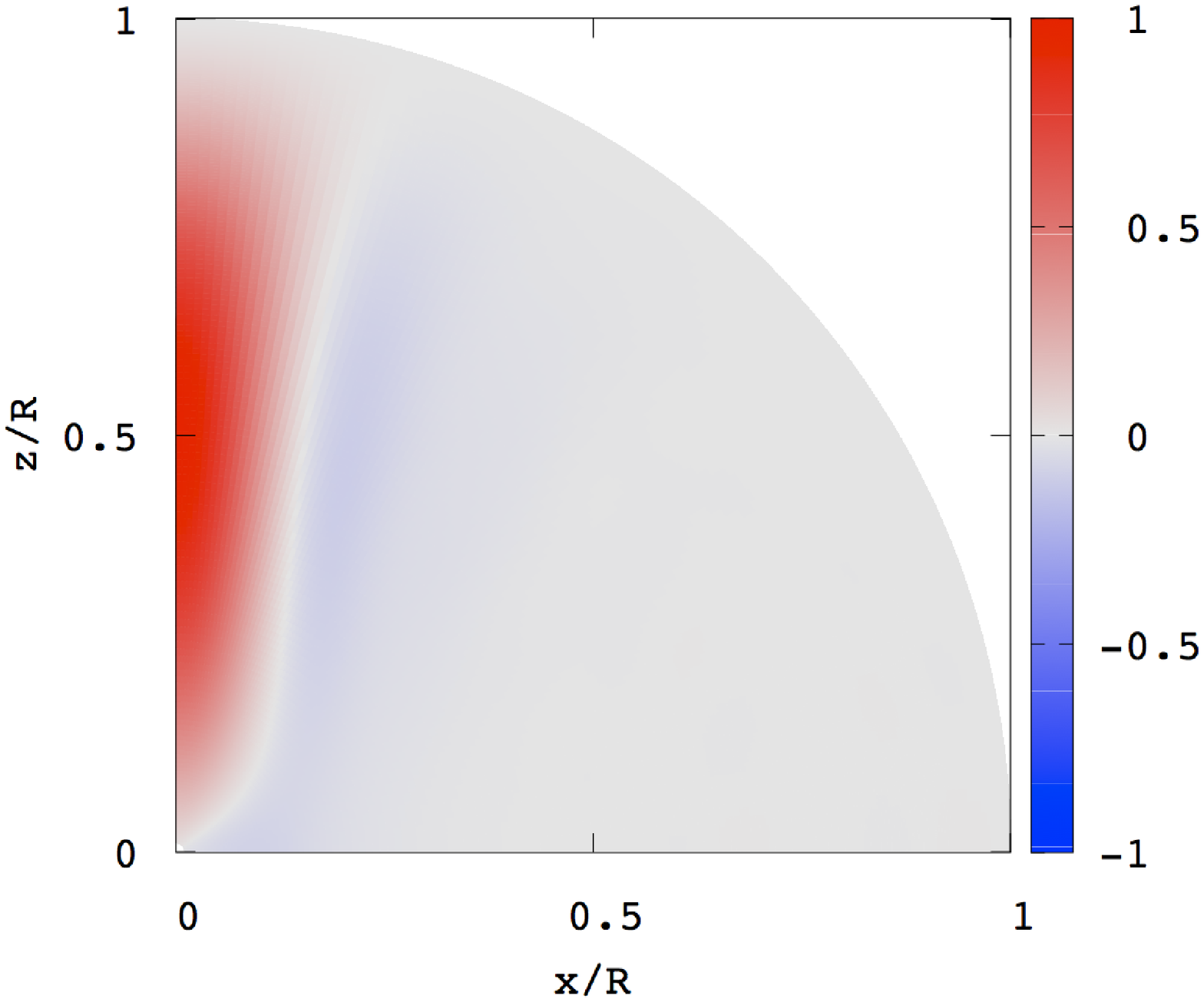}}
\hspace*{-1.41cm}
\resizebox{0.33\columnwidth}{!}{
\includegraphics{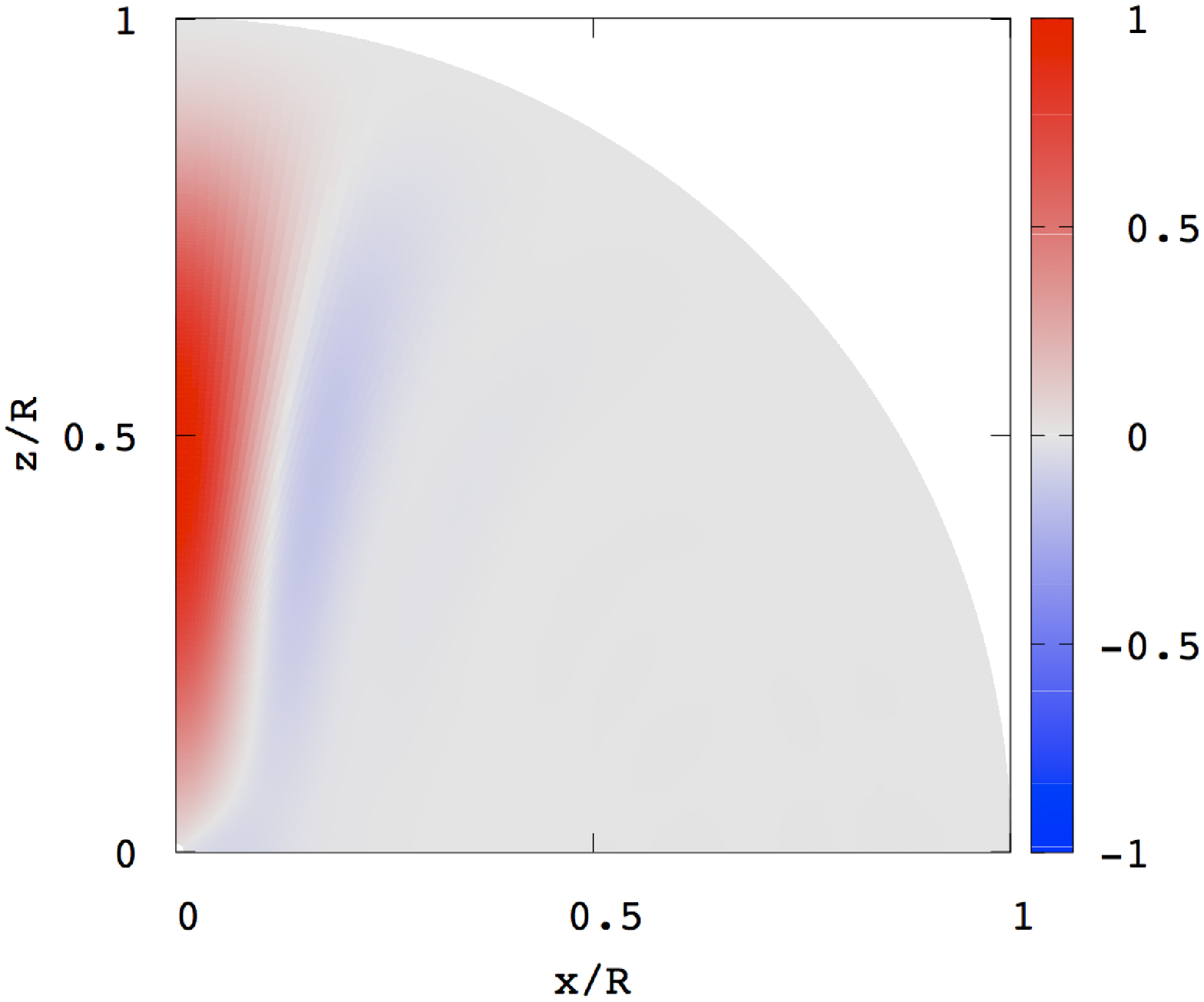}}
\end{center}
\vspace*{-1.5cm}
\begin{center}
\resizebox{0.33\columnwidth}{!}{
\includegraphics{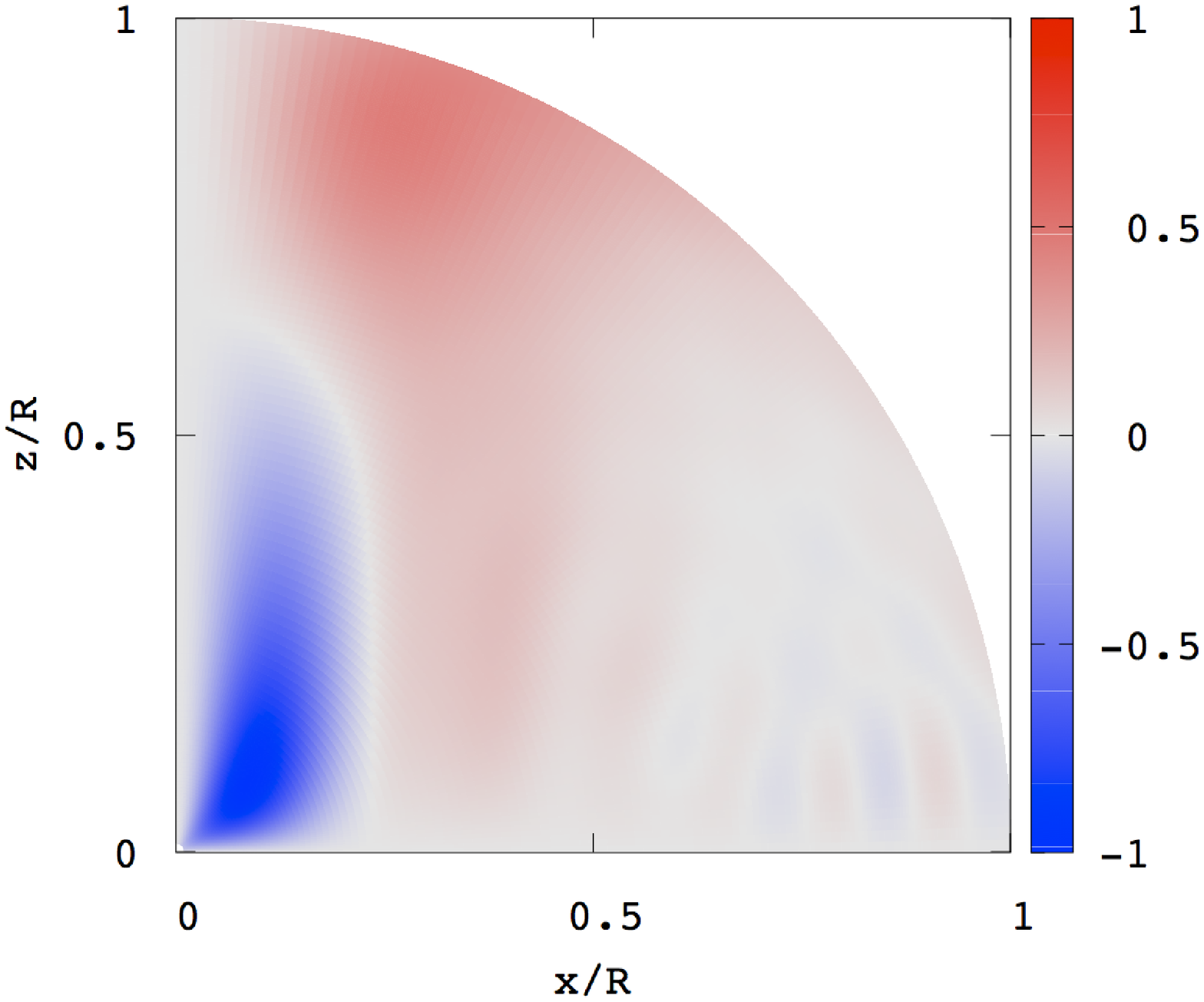}}
\hspace*{-1.41cm}
\resizebox{0.33\columnwidth}{!}{
\includegraphics{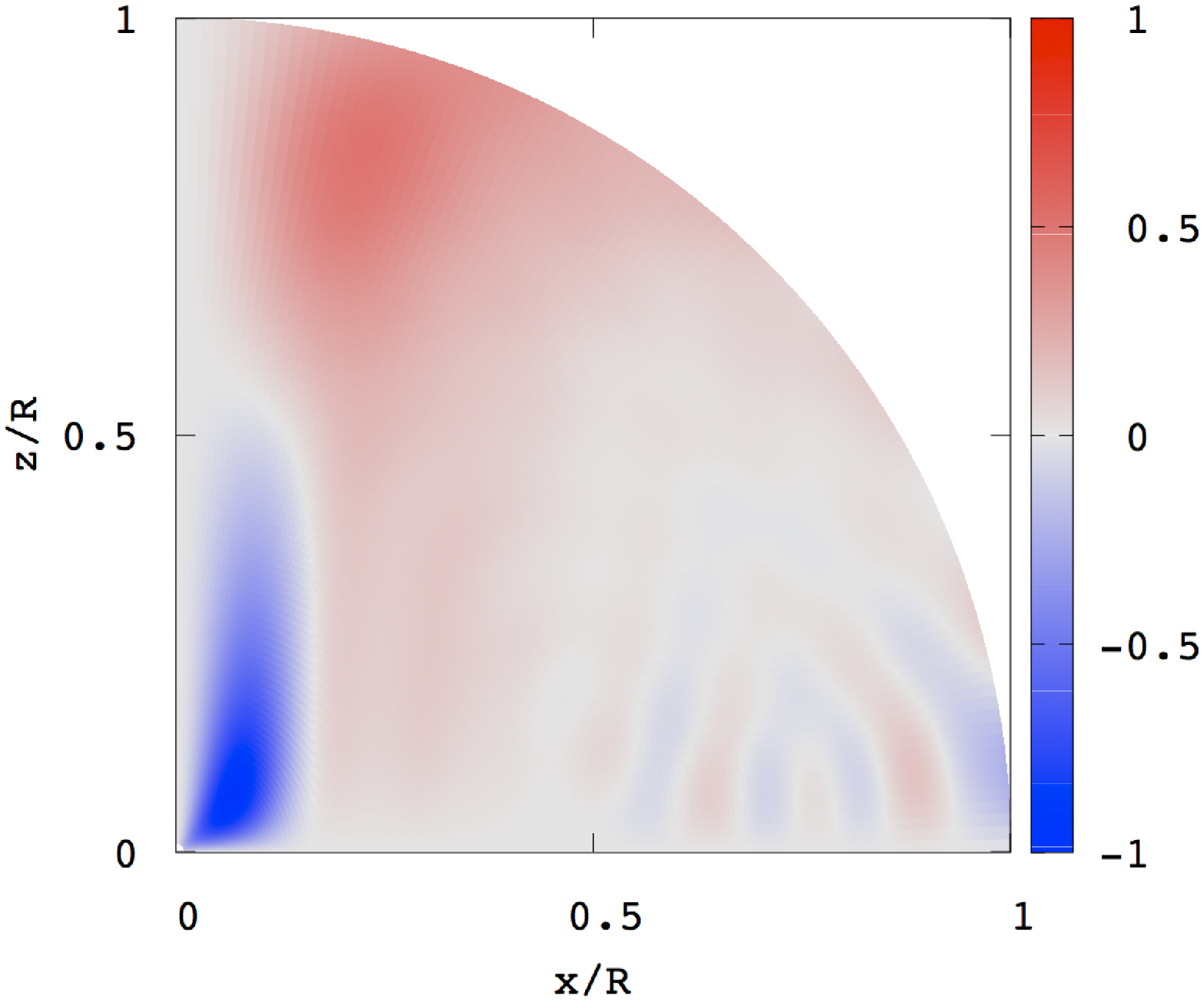}}
\hspace*{-1.41cm}
\resizebox{0.33\columnwidth}{!}{
\includegraphics{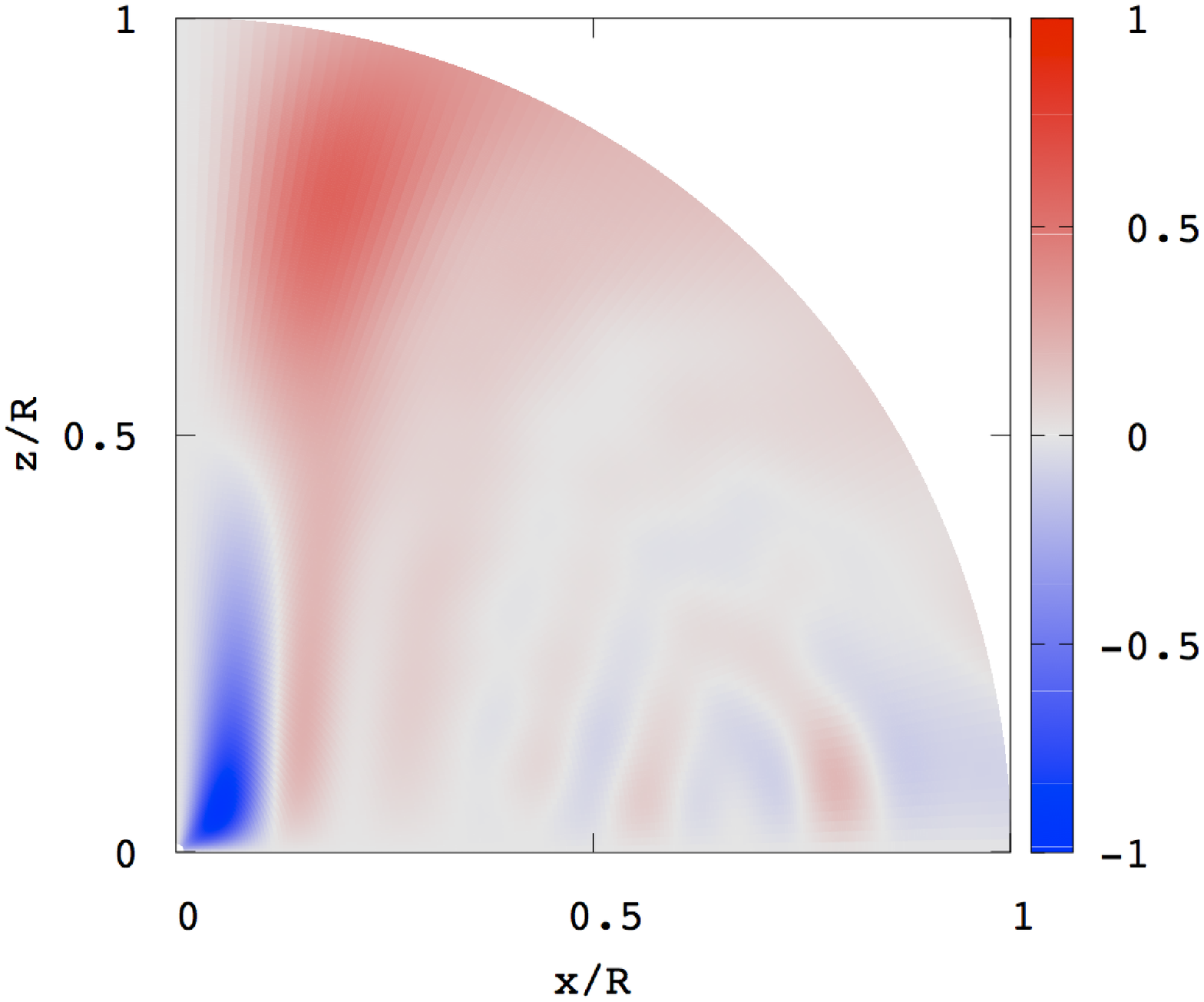}}
\end{center}
\vspace*{-1.3cm}
\caption{Wave patterns $\xi_r(x,z)$ (top panels) and $\xi_\theta(x,z)$ (bottom panels)
of the axisymmetric even magnetic modes of polytropes of the indices 
$n=0.5$, 1, and 1.5, from the left to right panels, where $B_S=10^{15}$G and $\gamma=0$ are assumed.
The normalized frequency $\bar\omega$ is $0.01154$, 0.01425, and 0.01835 for $n=0.5$, 1, and 1.5, respectively.
Note that the amplitudes of $\xi_r(x,z)$ and $\xi_\theta(x,z)$ are normalized by their maximum values.
}
\end{figure}

\begin{figure}
\begin{center}
\resizebox{0.33\columnwidth}{!}{
\includegraphics{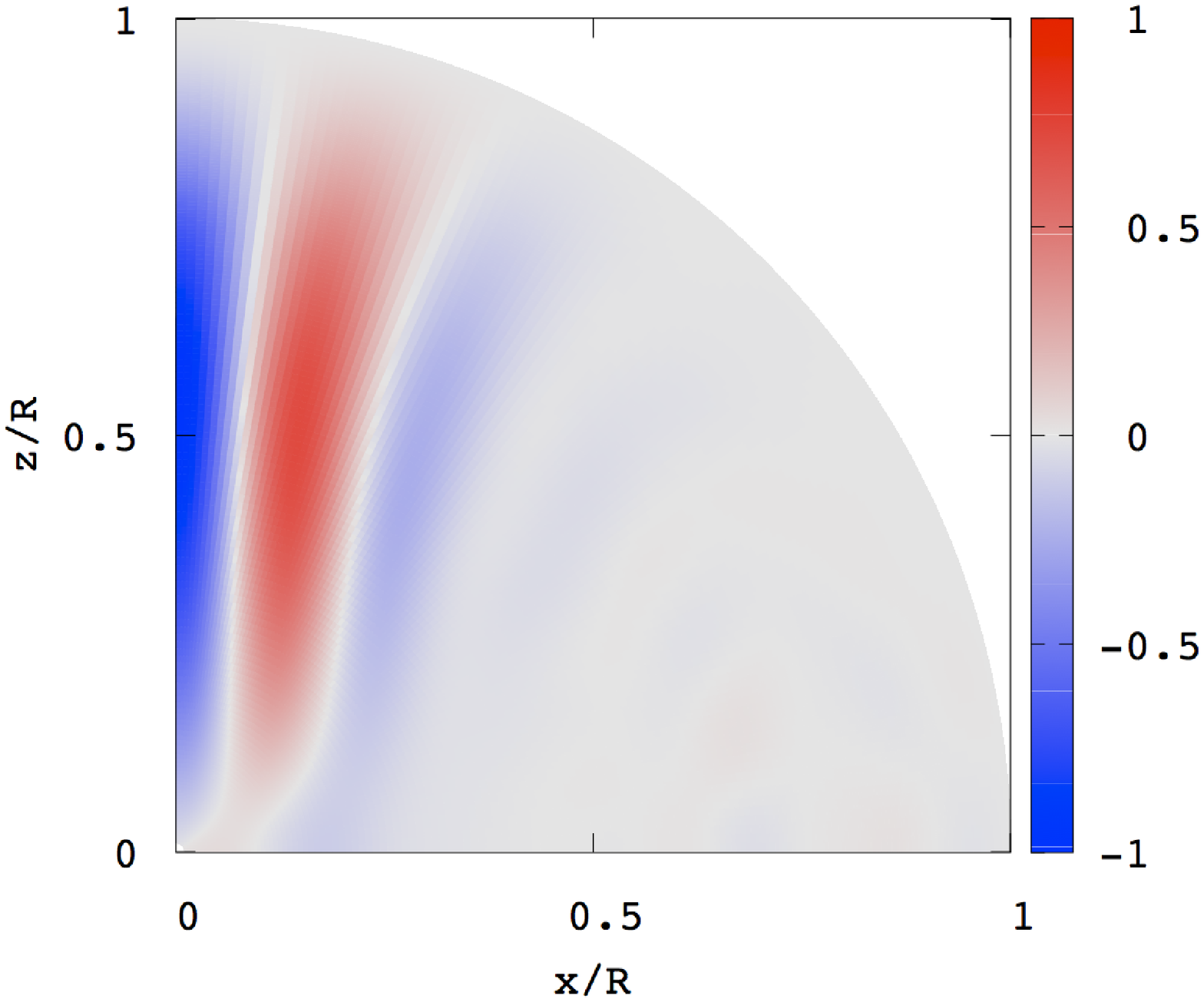}}
\hspace*{-1.41cm}
\resizebox{0.33\columnwidth}{!}{
\includegraphics{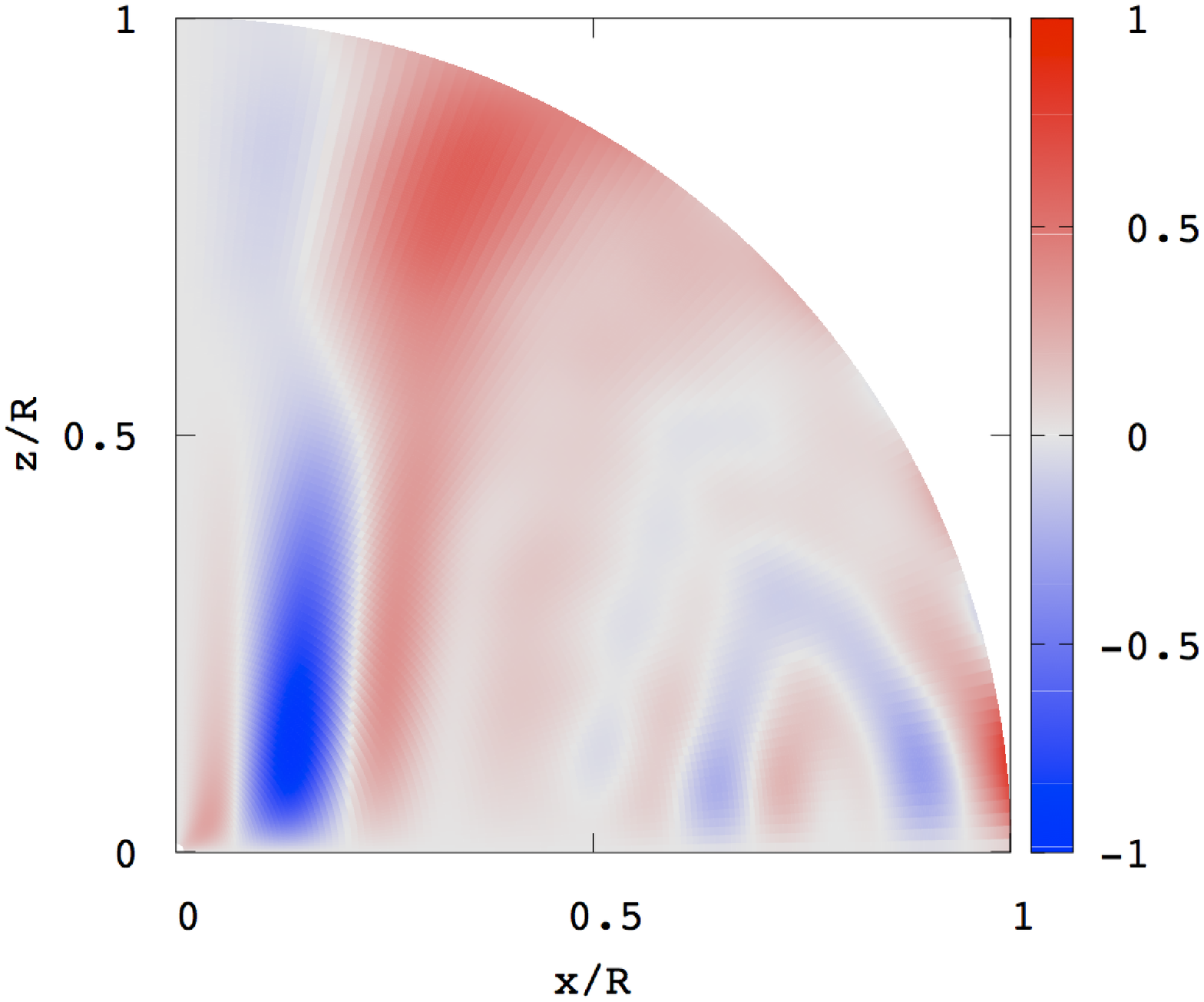}}
\vspace*{-1.3cm}
\end{center}
\caption{
Wave patterns $\xi_r(x,z)$ (left panel) and $\xi_\theta(x,z)$ (right panel)
of the axisymmetric even magnetic modes $\bar\omega=0.01400$ of the $n=1$ polytropes for $B_S=10^{15}$G and $\gamma=0$.
The amplitudes are normalized by their maximum values.}
\end{figure}

\begin{figure}
\begin{center}
\resizebox{0.33\columnwidth}{!}{
\includegraphics{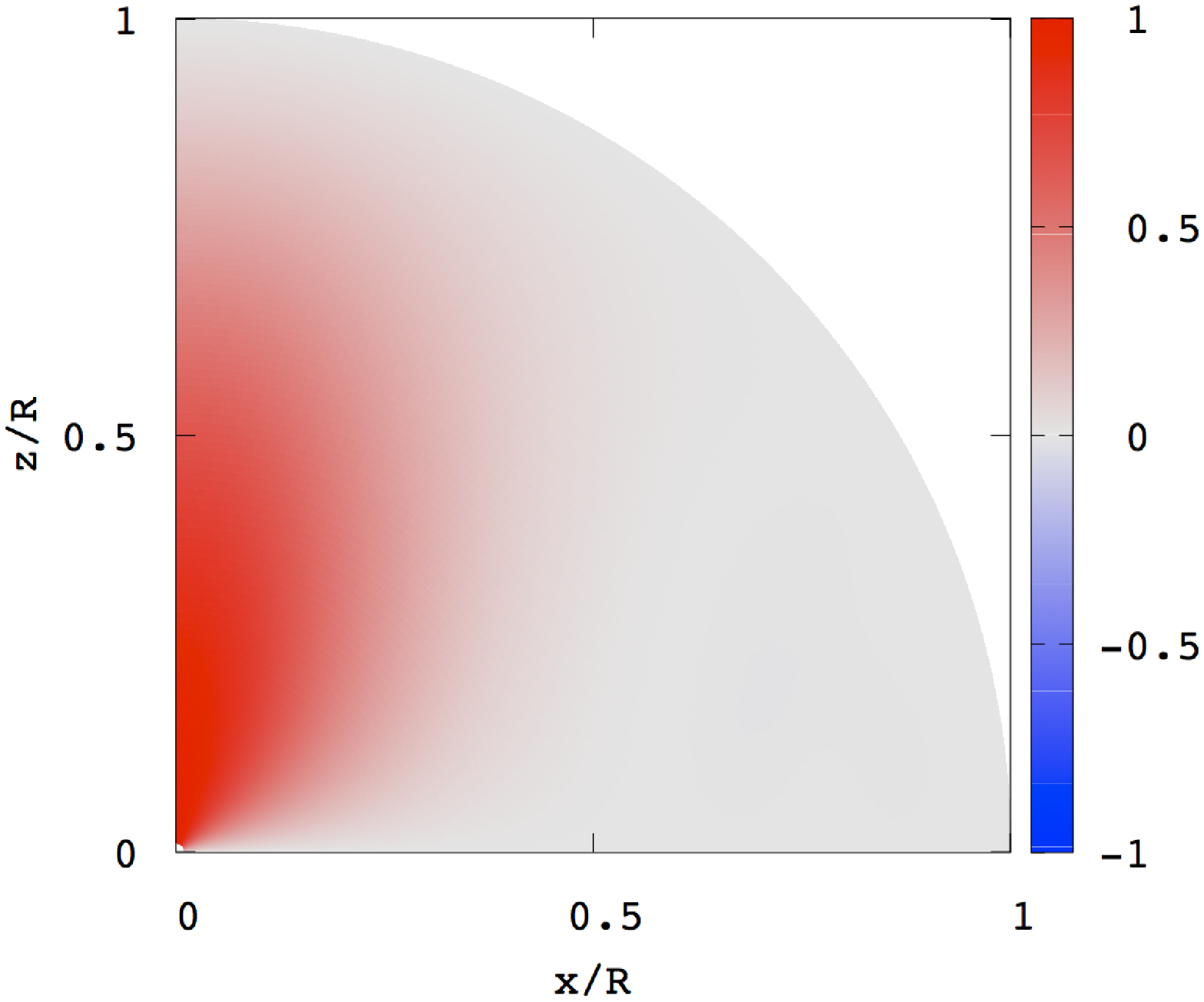}}
\hspace*{-1.41cm}
\resizebox{0.33\columnwidth}{!}{
\includegraphics{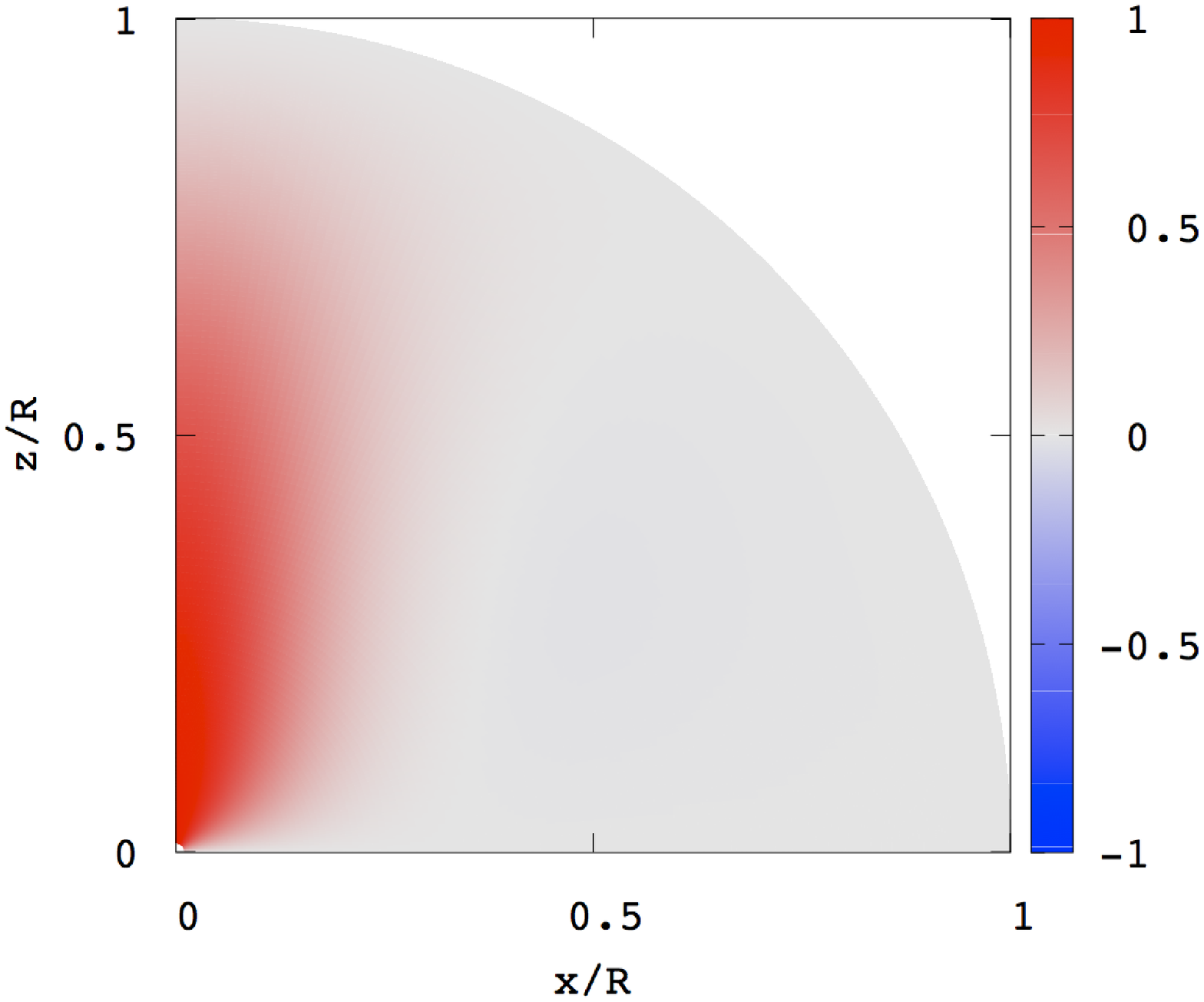}}
\hspace*{-1.41cm}
\resizebox{0.33\columnwidth}{!}{
\includegraphics{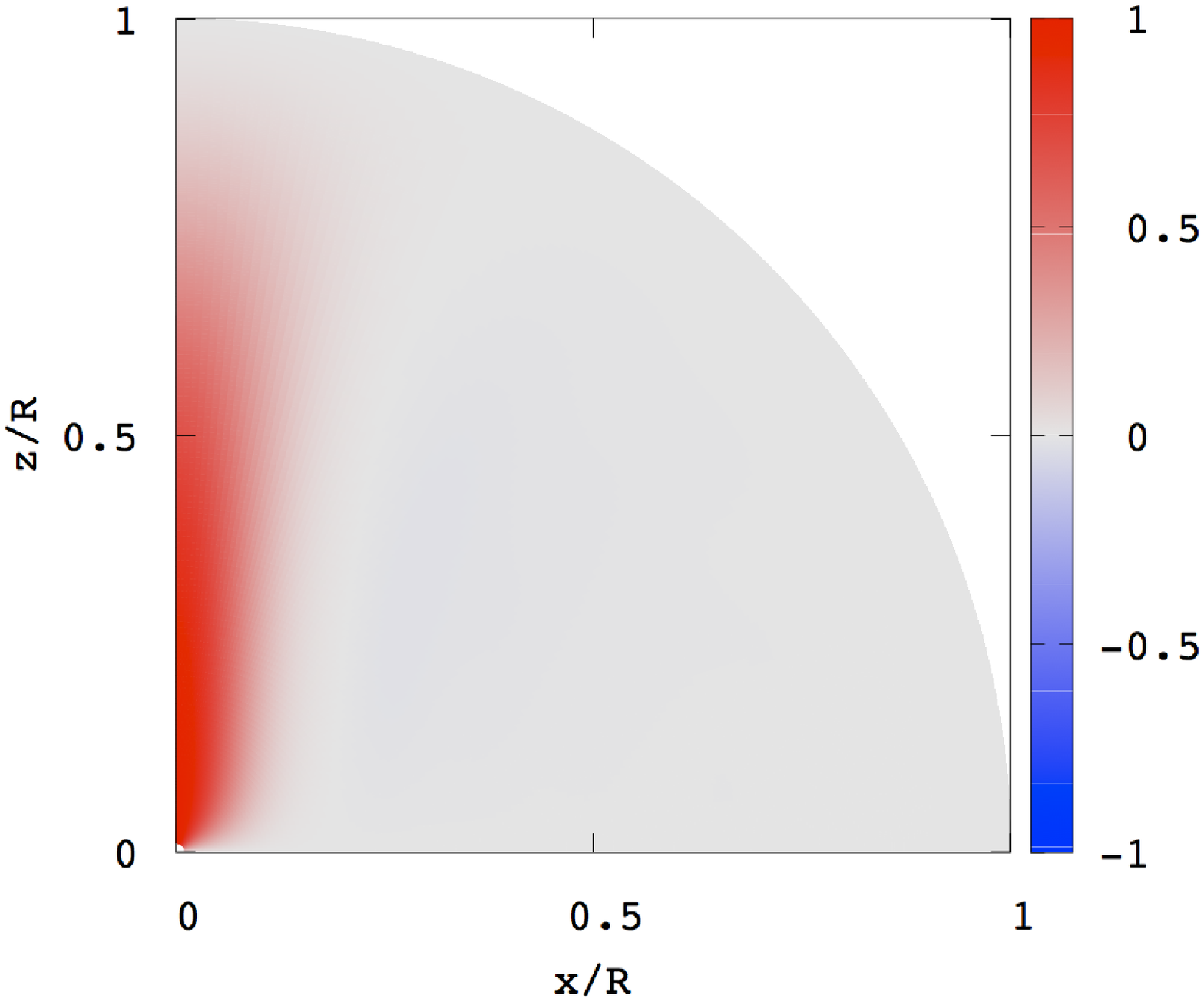}}
\end{center}
\vspace*{-1.5cm}
\begin{center}
\resizebox{0.33\columnwidth}{!}{
\includegraphics{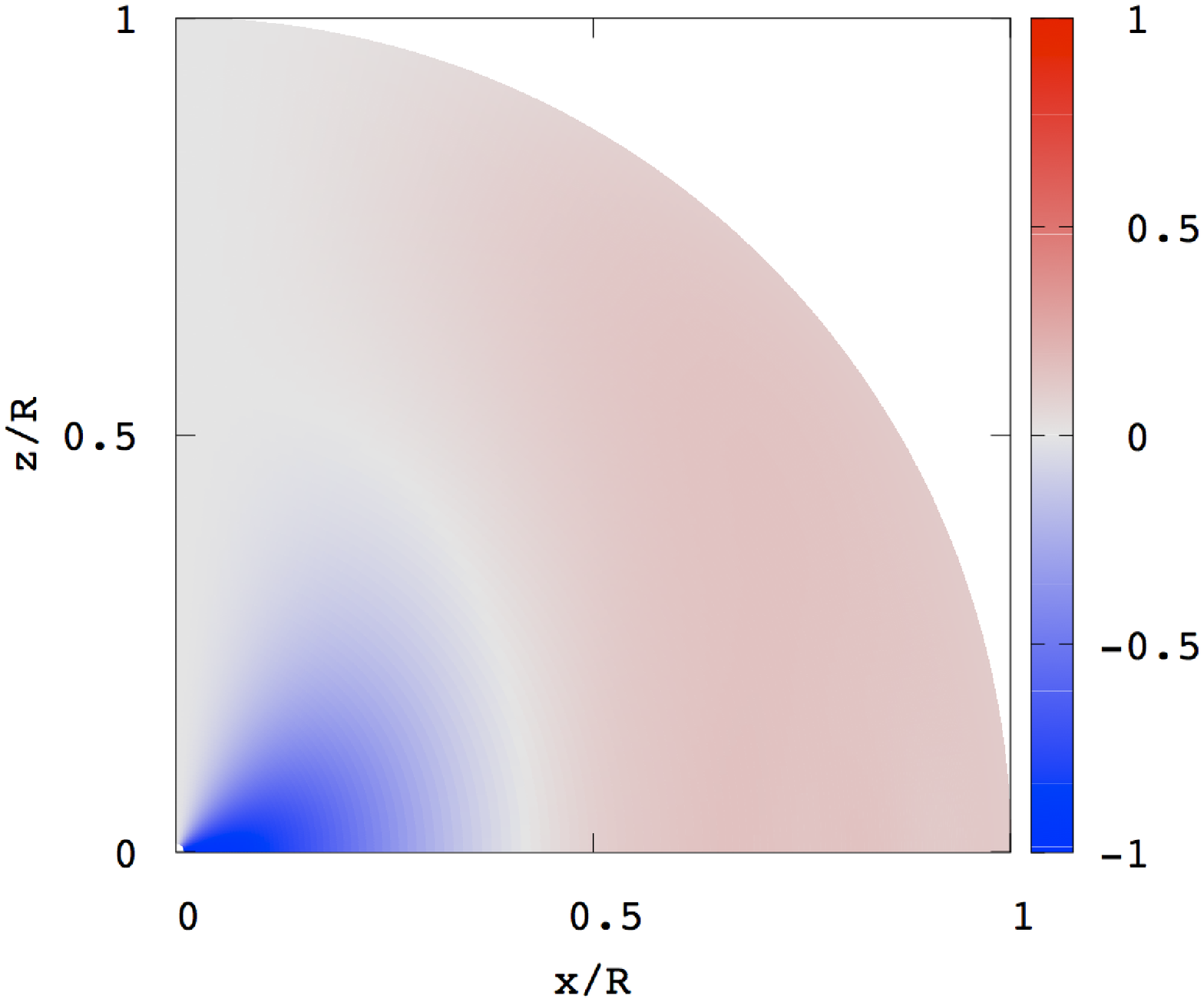}}
\hspace*{-1.41cm}
\resizebox{0.33\columnwidth}{!}{
\includegraphics{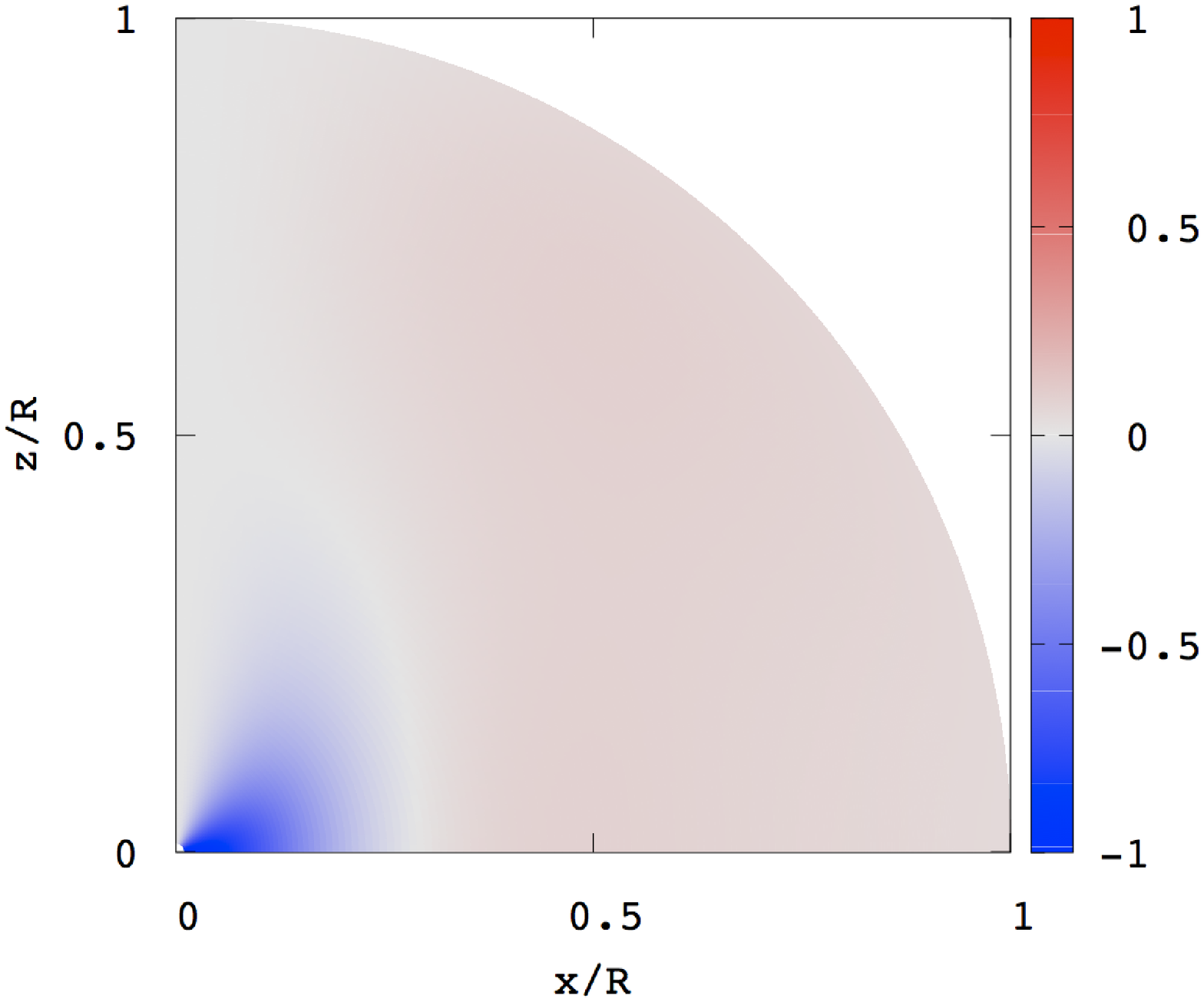}}
\hspace*{-1.41cm}
\resizebox{0.33\columnwidth}{!}{
\includegraphics{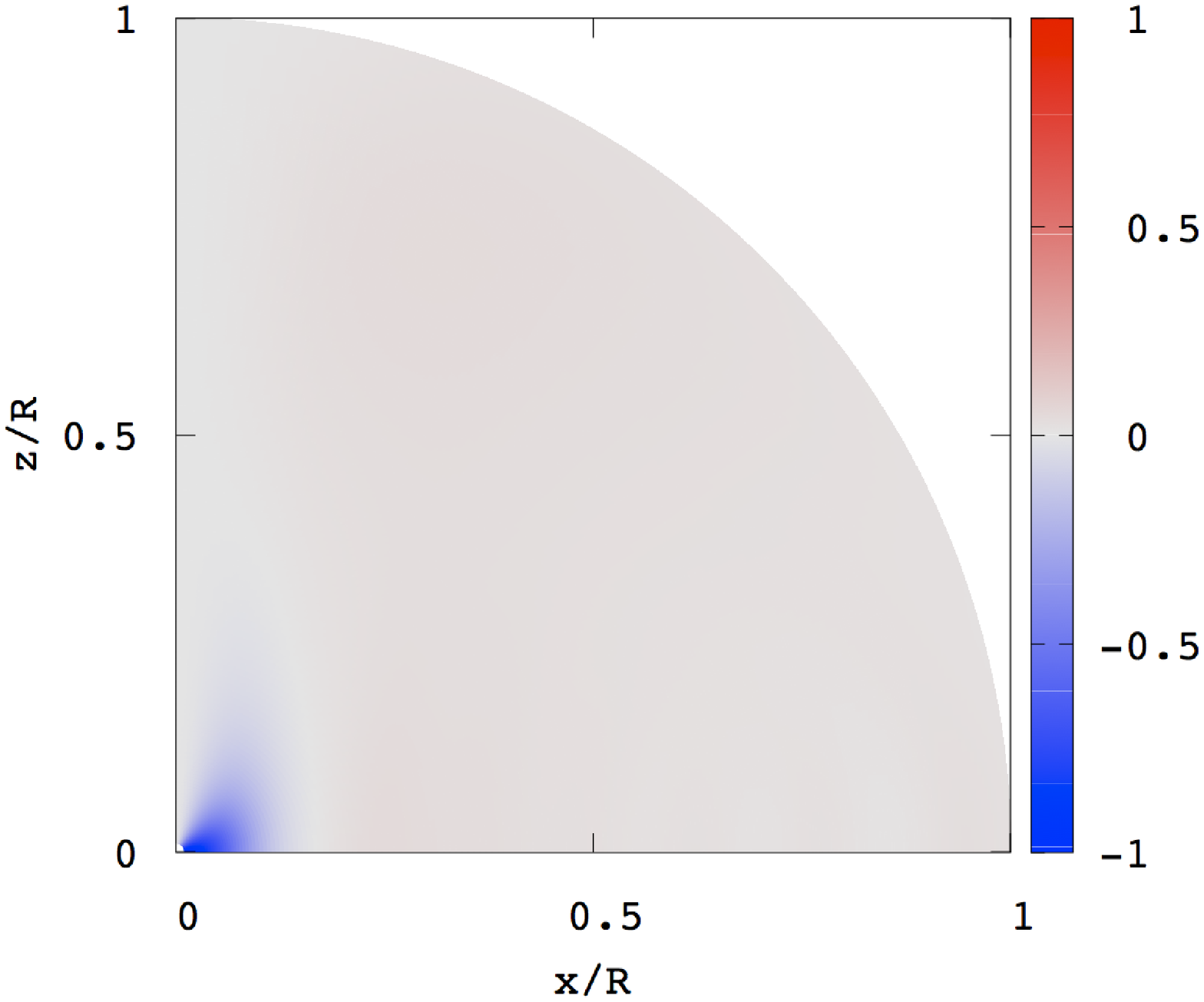}}
\end{center}
\vspace*{-1.3cm}
\caption{Same as Figure 3 but for the magnetic modes of odd parity.
The eigenfrequency $\bar\omega$ is 0.005510, 0.007268, and 0.01001 for
$n=0.5$, 1, and 1.5, respectively.
}
\end{figure}

We have examined the response of the frequency $\bar\omega$ of the magnetic modes to buoyancy by increasing $|\gamma|$ from $\gamma=0$
for $\gamma<0$ and we find the behavior of the frequency with increasing $|\gamma|$ is the same as
that found by Asai et al (2016) for non-axisymmetric spheroidal magnetic modes.
Note that we find no unstable magnetic modes for axisymmetric spheroidal oscillations.

\subsection{Modes Corresponding to Radial Acoustic Modes}

Using the Alfv\'en velocity given by $v_A=B_S/\sqrt{4\pi\rho_c}$ with $\rho_c$ being
the mass density at the stellar centre, we may define the lower limit to the frequency of Alfv\'en modes
as $\omega_{\rm LL}\equiv\pi v_A/R$ with $R$ being the radius of the star.
For $B_S=10^{15}$G, we have $\bar\omega_{\rm LL}=1.86\times10^{-3}$, $1.39\times10^{-3}$,
and $1.03\times10^{-3}$ for the polytropes of $n=0.5$, 1, and 1.5, respectively.
These small values of $\bar\omega_{\rm LL}$ suggest that Alfv\'en waves have very short wavelengths in the frequency region 
$\bar\omega\gtsim 1$ of acoustic modes, which makes it difficult to numerically correctly calculate
acoustic modes that are coupled with very short Alfv\'en waves.
Interestingly, however, we find a few acoustic modes corresponding to the radial ($l=0$) fundamental and first harmonic modes of $\bar\omega\gtsim 1$ for magnetized stars, although we find it very difficult to obtain acoustic modes corresponding to non-radial ($l\not=0$) $p$-modes.
The results are given in Table 2 where $\bar\omega$ is tabulated for two lowest radial order modes corresponding to 
the radial fundamental and first harmonic modes of polytropes of the indices $n=0.5$, 1, and 1.5 for
$B_S=10^{14}$G, $10^{15}$G, and $10^{16}$G.
As the table indicates the effect of the magnetic field of strength $B_S\ltsim 10^{16}$G on the high frequency modes
is quite minor.
For the field strength $B_S=10^{14}$G the frequencies $\bar\omega$ we obtain are practically those of radial pulsations obtained by ignoring the magnetic fields.
As $B_S$ increases, the frequency deviation from the radial pulsation modes increases 
but it is still quite small even for $B_S\sim 10^{16}$G.
This deviation is found larger for larger $n$ polytropes and for the radial first harmonic modes compared to
the radial fundamental modes. 
For a magnetic field as strong as $B_S\gtsim 10^{17}$G, however,
it becomes difficult to find well converged modes that correspond to the
radial modes of non-magnetized stars.
This is probably because the frequencies of magnetic modes of lowest radial order become comparable to
those of the radial fundamental and first harmonic modes.
The expansion coefficients $S_{l_1}$ and $x^2p'_{l_1}/\rho g r$ are plotted versus $x=r/R$ for the $n=1$ polytrope for $B_S=10^{14}$G. 
The functions $S_{l_1}$ and $x^2p'_{l_1}/\rho g r$ are almost indistinguishable from those of
radial pulsation modes.

\begin{figure}
\begin{center}
\resizebox{0.45\columnwidth}{!}{
\includegraphics{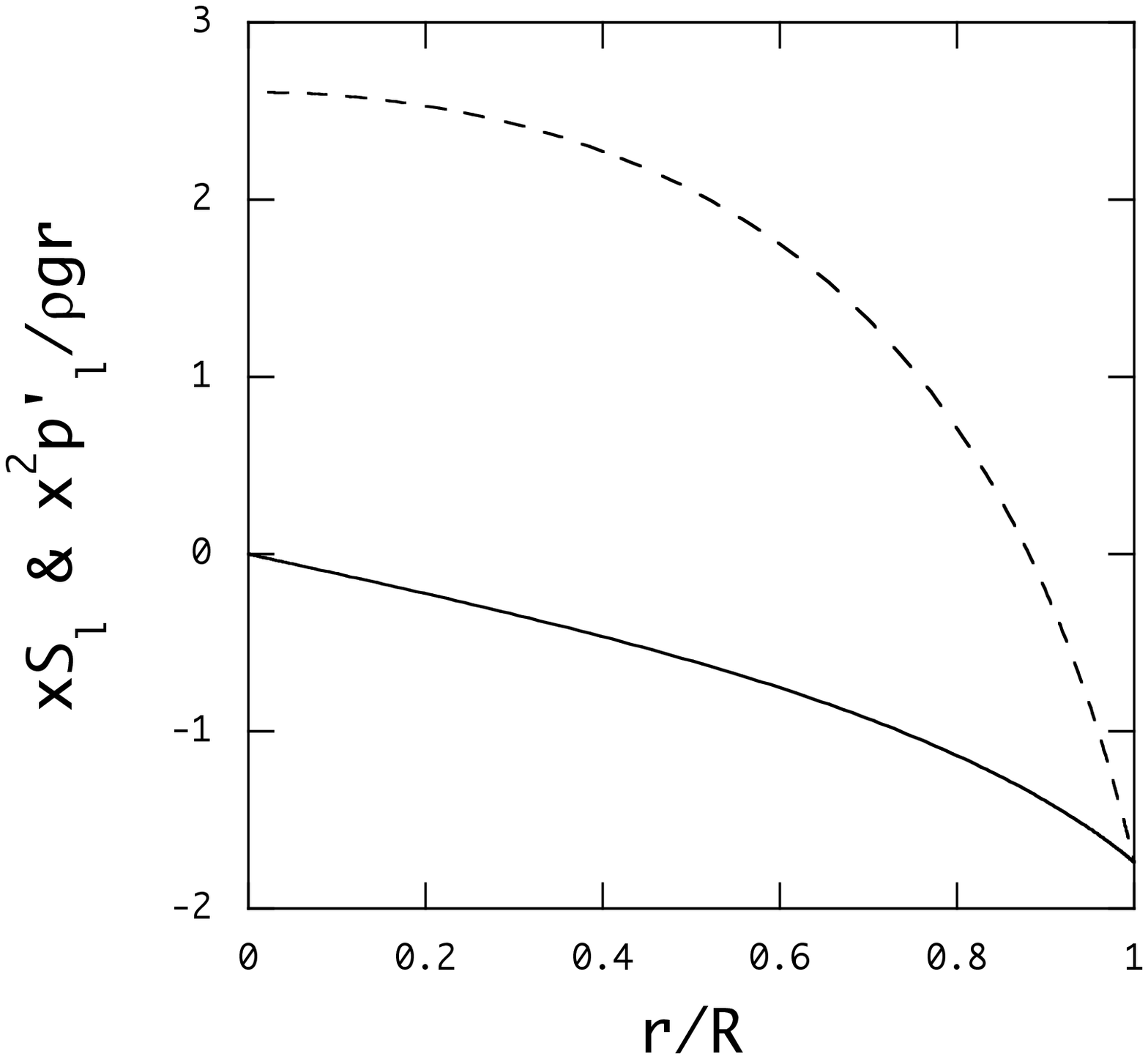}}
\resizebox{0.45\columnwidth}{!}{
\includegraphics{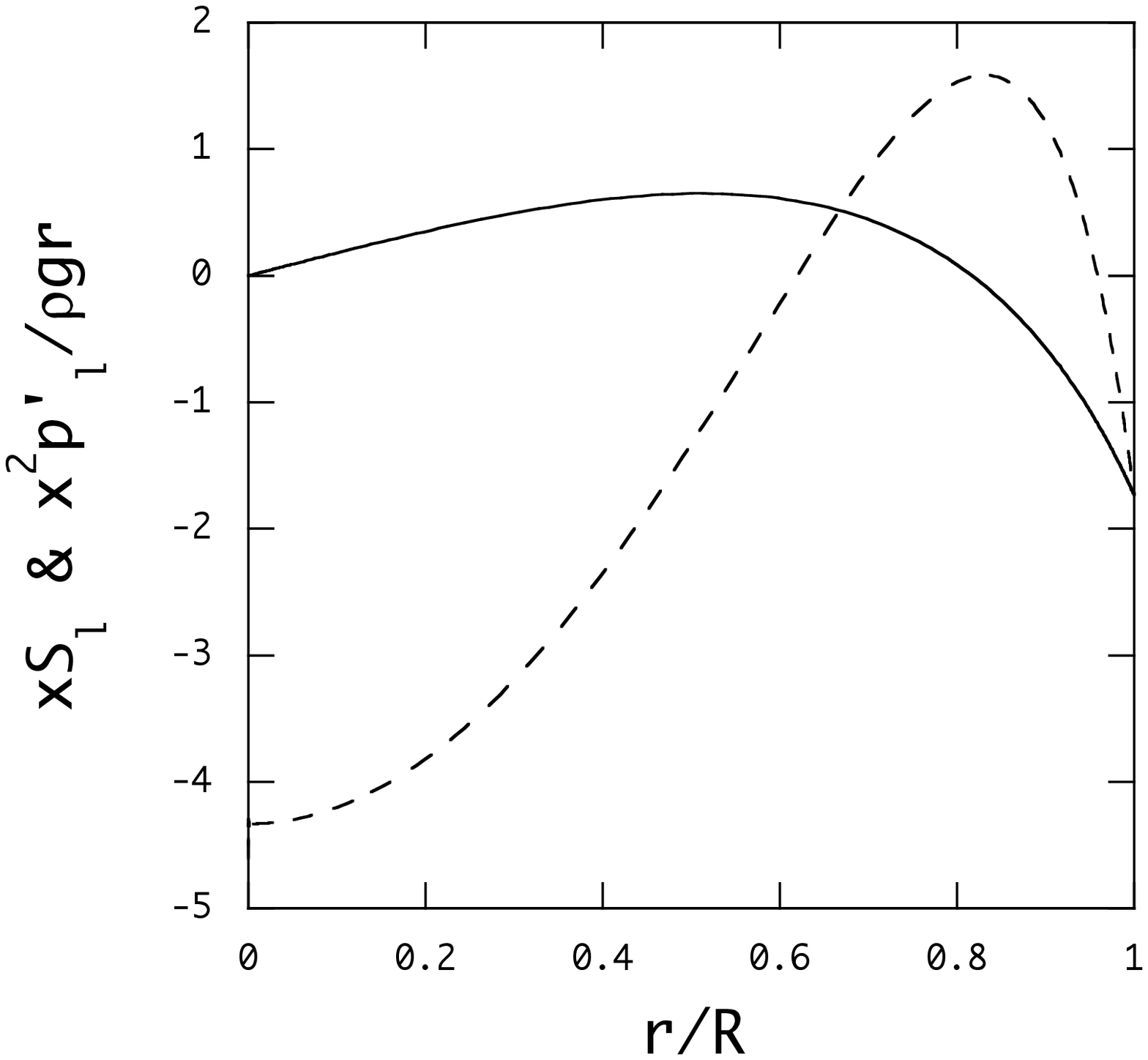}}
\end{center}
\caption{Expansion coefficients $S_{l_1}$ (solid line) and $x^2p'_{l_1}/\rho gr$ (dashed line) versus $x=r/R$ for
the modes corresponding to the radial fundamental (left panel) and first harmonic (right panel) modes
of the $n=1$ polytrope for $B_S=10^{14}$G where $M=1.4M_\odot$ and $R=10^6$cm.
The amplitude normalization is given by $b^H_{l'_1}=1$ at the surface.
}
\end{figure}

\begin{table*}
\begin{center}
\caption{Eigenfrequency $\bar{\omega}$ of acoustic modes corresponding to the radial fundamental and first harmonic modes 
of polytropes for the indices $n=0.5$, 1, and 1.5 for $\gamma=0$.}
\begin{tabular}{ccccccccc}
\hline
\multicolumn{3}{c}{$B_S=10^{14}$G}&\multicolumn{3}{c}{$B_S=10^{15}$G}&\multicolumn{3}{c}{$B_S=10^{16}$G} \\
\hline
$n=0.5$ & $n=1$ & $n=1.5$ & $n=0.5$ & $n=1$ & $n=1.5$ & $n=0.5$ & $n=1$ & $n=1.5$\\
\hline
3.0982 & 2.5930  & 2.4451  & 3.0982 & 2.5930  & 2.4451 & 3.0983 & 2.5933 & 2.4456\\
5.4026 & 4.3937  & 4.0237  & 5.4026 & 4.3937  & 4.0238 & 5.4036 & 4.3967 & 4.0310\\
\hline
\hline
\end{tabular}
\medskip
\end{center}
\end{table*}

\section{Conclusions}

We have computed axisymmetric spheroidal normal modes of polytropes magnetized with a poloidal magnetic field.
This paper may be regarded as an addition to a series of papers that discuss normal modes of magnetized stars (Lee 2005, 2007, 2008; Asai \& Lee 2014, Asai, Lee, \& Yoshida 2015, 2016).
Spheroidal and toroidal modes of stars magnetized with a poloidal field
are decoupled for axisymmetric oscillations of $m=0$, while they are coupled 
for non-axisymmetric oscillations of $m\not=0$ even for a purely poloidal magnetic field (e.g., Lee 2008).
In this paper, assuming axisymmetric spheroidal oscillations
we obtained magnetic modes of low frequency $\bar\omega\sim\bar\omega_{\rm LL}$
and modes of high frequency $\bar\omega\gtsim ~1$ that correspond to acoustic radial $l=0$ modes, where
the frequency of the former is approximately proportional to the field strength $B_S$ and that of the latter
is almost insensitive to $B_S$.
This result may be consistent with that obtained by Sotani \& Kokkotas (2009), who employed MHD numerical simulations for small amplitude oscillations.
As in the case of non-axisymmetric magnetic normal modes (Asai, Lee, \& Yoshida 2016),
as the frequency of the magnetic modes decreases
the wave patterns become complex having more nodal lines in the patterns and it becomes
more and more difficult to obtain well converged modes.
It is also to be noted that we could not obtain $g$-modes of magnetized stars for $\gamma<0$, the result of which
is the same as that by Asai et al (2016) for non-axisymmetric spheroidal oscillations.
Note also that we found no unstable magnetic modes with $\omega^2<0$ for axisymmetric modes
although Asai et al (2016) found unstable magnetic modes for non-axisymmetric ones (see also Lander \& Jones 2011).

As suggested by Table 1, the frequencies of axisymmetric spheroidal magnetic modes computed for $B_S\sim 10^{15}$G
are consistent with low frequency ($\sim 30$Hz) QPOs detected for SGR 1806-20 and SGR 1900+14.
However, it is obvious that all the identified QPOs for the SGRs cannot be explained in terms of 
spheroidal magnetic normal modes alone.

Using MHD simulations for mixed poloidal and toroidal magnetic fields, for example,
Colaiuda \& Kokkotas (2012) indicated that the frequency spectra of
toroidal modes will be significantly modified and lose continuum character.
Since we do not have any reliable knowledges concerning magnetic field configuration 
in the interior of neutron stars,
it is useful to study normal modes of magnetized stars for more general magnetic field configurations that consist of
both poloidal and toroidal fields.
As a first step toward such normal mode analyses we may add a weak toroidal field component to the dominant poloidal one 
in order to examine how the normal modes respond to such a weak toroidal component.


\begin{thebibliography}{99}
%\bibitem[\protect\citeauthoryear{}{}] {} Arras P., Cumming A.,
%  Thompson C., 2004, ApJ, 608, L49
\bibitem[\protect\citeauthoryear{AsaiLee}{2014}] {} Asai H., Lee U.,
  2014, ApJ, 790, 66
\bibitem[\protect\citeauthoryear{}{}] {} Asai H., Lee U., Yoshida S.,
  2015, MNRAS 449, 3620
\bibitem[\protect\citeauthoryear{}{}] {} Asai H., Lee U., Yoshida S.,
  2016, MNRAS 449, 3620
%\bibitem[\protect\citeauthoryear{}{}] {} Braithwaite J., Spruit H. C.,
%  2004, Nature, 431, 819
%\bibitem[]{} Braithwaite, J., 2007, A\&A, 469, 275
\bibitem[\protect\citeauthoryear{Akgunetal}{2013}]{} Akg\"un T., Reisenegger A., Mastrano A., Marchant P., 2013, MNRAS, 433, 2445
\bibitem[\protect\citeauthoryear{Cerd$\acute{{\rm{a}}}$-Dur$\acute{{\rm{a}}}$n
  et al}{2009}]{b22} Cerd$\acute{{\rm{a}}}$-Dur$\acute{{\rm{a}}}$n P.,
Stergioulas N., Font J. A., 2009, MNRAS, 397, 1607 %axisymmetric toroidal for a poloidal field without a crust
%\bibitem[\protect\citeauthoryear{}{}] {} Ciolfi R., Ferrari V.,
%  Gualtieri L., Pons J. A., 2009, MNRAS, 397, 913
%\bibitem[\protect\citeauthoryear{}{}] {} Ciolfi R., Rezzolla L., 2012,
%  ApJ, 760, 1
\bibitem[\protect\citeauthoryear{}{}] {} Colaiuda A., Ferrari V.,
  Gualtieri L., Pons J. A., 2008, MNRAS, 385, 2080  
\bibitem[\protect\citeauthoryear{Colaiuda \& Kokkotas}{2011}]{b2} Colaiuda A., Kokkotas K. D., 2011, MNRAS, 414, 3014
%axisymmetric torsional oscillations for purely poloidal field
\bibitem[\protect\citeauthoryear{}{}] {} Colaiuda A., Kokkotas K. D., 2012,
  MNRAS, 423, 818 %coupled axial and polar axisymmetric oscillations for poloidal + toroidal fields
%\bibitem[]{} Glampedakis K., Samuelsson L., Andersson N., 2006, MNRAS,
%  371, L74
%\bibitem[]{} Glampedakis K., Andersson N., Samuelsson L., 2011, MNRAS,
%  410, 805  %formulation
\bibitem[\protect\citeauthoryear{Gabler et al.}{2011}]{b5} Gabler M.,
  Cerd$\acute{{\rm{a}}}$-Dur$\acute{{\rm{a}}}$n P., Font J. A.,
  M$\ddot{{\rm{u}}}$ller E., Stergioulas N., 2011, MNRAS, 410, L37 %axisymmetric torsional oscillations
\bibitem[\protect\citeauthoryear{Gabler et al.}{2012}]{b5} Gabler M., Cerd$\acute{{\rm{a}}}$-Dur$\acute{{\rm{a}}}$n P., Stergioulas N., Font J. A.,
  M$\ddot{{\rm{u}}}$ller E., 2012, MNRAS, 421, 2054 %axisymmetric toroidal 
\bibitem[\protect\citeauthoryear{Gabler et al.}{2013a}]{b5} Gabler M.,
  Cerd$\acute{{\rm{a}}}$-Dur$\acute{{\rm{a}}}$n P., Font J. A.,
  M$\ddot{{\rm{u}}}$ller E., Stergioulas N., 2013, MNRAS, 430, 1811 %axisymmetric toroidal 
%\bibitem[\protect\citeauthoryear{Gabler et al.}{2013b}]{b5} Gabler M.,
%  Cerd$\acute{{\rm{a}}}$-Dur$\acute{{\rm{a}}}$n P., Stergioulas N., Font J. A.,
%  M$\ddot{{\rm{u}}}$ller E., 2013, PhRvL, 111
\bibitem[\protect\citeauthoryear{}{}]{} Herbrik M., Kokkotas K.D., 2017, MNRAS, 466, 1330
\bibitem[\protect\citeauthoryear{}{}] {} Israel G.,
  Belloni T., Stella L., Rephaeli Y., Gruber D. E., Casella P.,
  Dall'Osso S., Rea N., Persic M., Rothschild R. E., 2005, ApJ, 628,
  L53
\bibitem[]{} Lander S. K., Jones D.I., Passamonti A., 2010, MNRAS,
  405, 318  %mixed poloidal and toroidal oscillations of rotating neutron stars with toroidal magnetic fields
\bibitem[\protect\citeauthoryear{Lander \& Jones}{2011}]{b1} Lander
  S. K., Jones D. I., 2011, MNRAS, 412, 1730 %non-axisymmetric oscillations of rotating neutron stars with poloidal fields 
%\bibitem[\protect\citeauthoryear{}{}] {} Lasky P. D., Zink B.,
%  Kokkotas K. D., Glampedakis K., 2011, ApJ, 735, L20
\bibitem[\protect\citeauthoryear{}{}] {} Lee U., 2005, MNRAS 357, 97
\bibitem[\protect\citeauthoryear{}{}] {} Lee U., 2007, MNRAS 374, 1015
\bibitem[]{} Lee U., 2008, MNRAS, 385, 2069
%\bibitem[]{} Lee U., 2010, MNRAS, 405, 1444
\bibitem[]{} Levin Y., 2006, MNRAS, 368, L35
\bibitem[]{} Levin Y., 2007, MNRAS, 377, 159
\bibitem[]{} Markey, P., Tayler, R. J., 1973, MNRAS, 163, 77
\bibitem[]{} Markey, P., Tayler, R. J., 1974, MNRAS, 168, 505
\bibitem[]{} Passamonti A., Lander S.K., 2013, MNRAS, 429, 767 %non-axisymmetric oscillations for poloidal and toroidal fields
\bibitem[]{} Passamonti A., Lander S.K., 2014, MNRAS, 438, 156 %axisymmetric toroidal oscillations for purely poloidal field
\bibitem[\protect\citeauthoryear{}{}] {} Sotani H., Kokkotas K. D.,
  Stergioulas N., 2008, MNRAS 385, L5
\bibitem[\protect\citeauthoryear{}{}] {} Sotani H., Colaiuda A.,
  Kokkotas K. D., 2008, MNRAS 385, 2161 %axisymmetric ? torsional oscillation of the crust with a poloidal field
\bibitem[\protect\citeauthoryear{}{}] {} Sotani H., Kokkotas K. D.,
  2009, MNRAS 395, 1163 %axisymmetric polar oscillations for poloidal fields
\bibitem[\protect\citeauthoryear{}{}] {} Strohmayer T. E., Watts
  A. L., 2005, ApJ, 632, L111
\bibitem[\protect\citeauthoryear{}{}] {} Strohmayer T. E., Watts
  A. L., 2006, ApJ, 653, 593
\bibitem[\protect\citeauthoryear{}{}] {} Tayler R.J., 1973, MNRAS, 161, 365
%\bibitem[]{} van Assche, W.,  Goossens, M.,  Tayler, R. J., 1982, A\&A, 109, 166
\bibitem[]{} van Hoven M.B., Levin Y., 2011, MNRAS, 410, 1036
\bibitem[]{} van Hoven M.B., Levin Y., 2012, MNRAS, 420, 3035
\bibitem[\protect\citeauthoryear{}{}] {} Watts A.L., 2011, arXiv:1111.0514v1
\bibitem[\protect\citeauthoryear{}{}] {} Watts A. L., Strohmayer T. E., 2006, ApJ, 637, L117

%\bibitem[\protect\citeauthoryear{}{}] {} Yoshida S., Eriguchi Y.,
%  2006, ApJS, 120, 353
%\bibitem[\protect\citeauthoryear{}{}] {} Yoshida S., Yoshida S., Eriguchi Y.,
%  2006, ApJ, 651, 462 
\end{thebibliography}
\end{document}